\documentclass[twocolumn]{autart}     

\usepackage[utf8]{inputenc}
\usepackage[T1]{fontenc}
\usepackage{graphicx}
\usepackage{subcaption}
\usepackage{float}

\usepackage{amsmath}
\usepackage{amssymb}
\usepackage{mathtools}
\usepackage{empheq}

\usepackage{xcolor}
\usepackage{ifthen}

\usepackage[table]{xcolor}
\usepackage{multirow}
\usepackage{enumitem}

\usepackage{nicematrix}
\usepackage{tikz}
\usetikzlibrary{automata, positioning, arrows}
\usetikzlibrary{calc}
\usetikzlibrary{fit,tikzmark}
\usetikzlibrary{backgrounds}
\usetikzlibrary{arrows.meta}
\usepackage{forest}
\usepackage{arydshln}

\usepackage{xcolor}
\usepackage{bm}
\usepackage{stackrel}

\usepackage[bookmarks=false]{hyperref}

\usepackage[authoryear]{natbib}

\edef\endfrontmatter{%
  \unexpanded\expandafter{\endfrontmatter}
  \noexpand\endNoHyper 
}

\usepackage{wrapfig}

\newcommand{\blue}[1]{}

\newcommand{\vect}[1]{\boldsymbol{#1}} 

\newcommand{\R}{\mathbb{R}}
\newcommand{\N}{\mathbb{N}}

\newcommand{\Z}{\mathbb{Z}}

\newcommand{\defEqual}{\coloneqq}
\newcommand{\argmin}{\operatorname{argmin}}

\newcommand{\myemptyset}{\varnothing}

\newtheorem{theorem}{Theorem}
\newtheorem{lemma}{Lemma}
\newtheorem{corollary}{Corollary}
\newtheorem{proposition}{Proposition}
\newtheorem{definition}{Definition}
\newtheorem{example}{Example}

\renewcommand{\pf}{\textbf{Proof.}}
\renewcommand{\qed}{$\blacksquare$}

\newcommand{\set}[1]{\mathcal{#1}}
\newcommand{\Sys}{\mathcal{S}}
\newcommand{\Cont}{\mathcal{C}}
\newcommand{\seq}[1]{\boldsymbol{#1}} 
\newcommand{\B}{\set{B}_{\times}} %

\newcommand{\T}{\operatorname{T}} 
\newcommand{\ASR}{\operatorname{ASR}} 
\newcommand{\FRR}{\operatorname{FRR}} 
\newcommand{\MCR}{\operatorname{MCR}} 
 
\newcommand{\ASRB}{\operatorname{PSR}} 
 
\newcommand{\ASRBB}{\operatorname{FAR}}

\newcommand{\length}[1]{[0;#1[} 

\definecolor{strongColor}{RGB}{215,215,215}

\allowdisplaybreaks[1]
\begin{document}

\begin{frontmatter}

\title{Characterizing simulation relations through control architectures in abstraction-based control\thanksref{footnoteinfo}} 

\thanks[footnoteinfo]{
This paper was not presented at any IFAC meeting. Corresponding author J.~Calbert. 
}

\author[UCL]{Julien Calbert}\ead{julien.calbert@uclouvain.be},    
\author[L2S]{Antoine Girard}\ead{antoine.girard@centralesupelec.fr},  
\author[UCL]{Rapha\"{e}l M. Jungers}\ead{raphael.jungers@uclouvain.be}               
\address[UCL]{ICTEAM Institute, Department of Mathematical Engineering, UCLouvain, B-1348, Louvain-la-Neuve, Belgium}  
\address[L2S]{Universit\'{e} Paris-Saclay, CNRS, CentraleSup\'{e}lec, Laboratoire des signaux et syst\`{e}mes, 91190, Gif-sur-Yvette, France}


\begin{keyword}                           
abstraction; symbolic model; simulation relations.               
\end{keyword}                             

\begin{abstract} 

Abstraction-based control design is a promising approach for ensuring safety-critical control of complex cyber-physical systems. A key aspect of this methodology is the relation between the original and abstract systems, which ensures that the abstract controller can be transformed into a valid controller for the original system through a concretization procedure.
In this paper, we provide a comprehensive and systematic framework that characterizes various simulation relations, through their associated concretization procedures. We introduce the concept of \emph{interfaced system}, which universally enables a feedback refinement relation with the abstract system. This interfaced system encapsulates the specific characteristics of each simulation relation within an \emph{interface}, enabling a plug-and-play control architecture.
Our results demonstrate that the existence of a particular simulation relation between the concrete and abstract systems is equivalent to the implementability of a specific control architecture, which depends on the considered simulation relation.
This allows us to introduce new types of relations, and to establish the advantages and drawbacks of different relations, which we exhibit through detailed examples.

\end{abstract}
\end{frontmatter}

\section{Introduction}\label{sec:intro}

\begin{figure}
    \centering
    \includegraphics[width=0.44\textwidth]{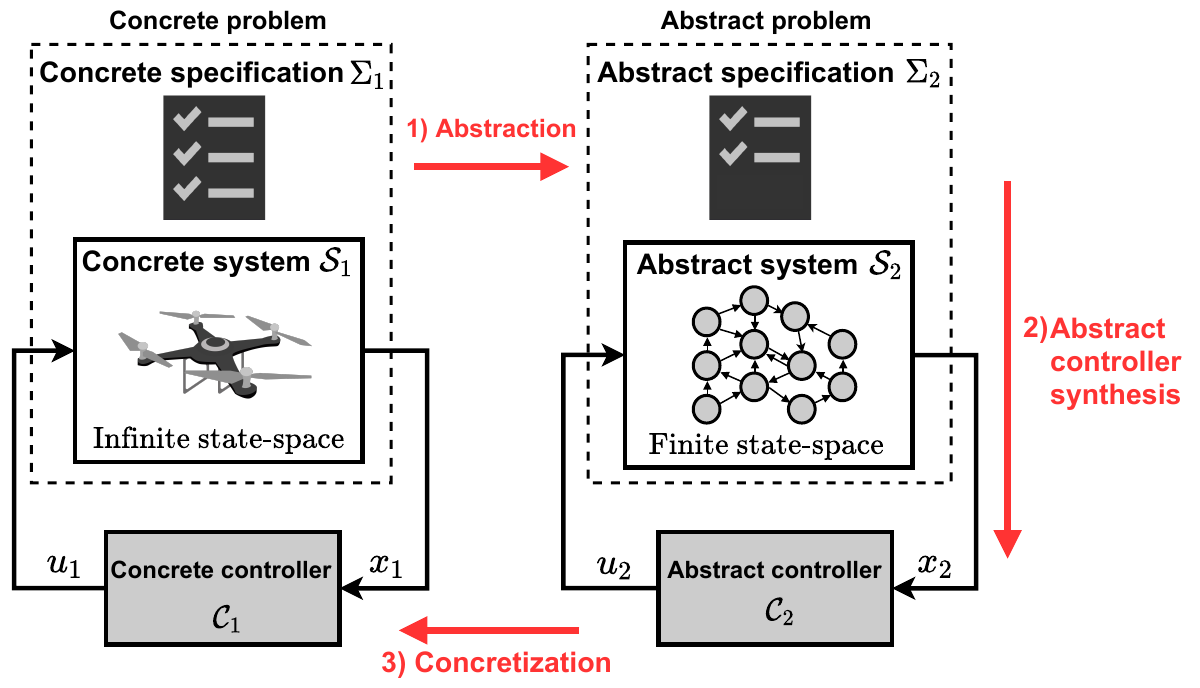}
    \caption{The three steps of abstraction-based control. 
    Our work focuses on the connection between the type of relation between $\Sys_1$ and $\Sys_2$ (step 1) and the concretization procedure (step 3). We show that one can tailor the abstraction relation depending on the required properties of the concretization step and of the finally obtained concrete controller, independently of the abstract controller synthesis step (step 2).
    }
    \label{fig:abstraction-based-control}%
\end{figure}

Abstraction-based control design is a thriving and promising proposal for safety-critical control of complex cyber-physical systems~\citep{alur2015principles,kim2012cyber}.
It stems from the idea of synthesizing a correct-by-construction controller through a systematic three-step procedure illustrated in Figure~\ref{fig:abstraction-based-control}. 
First, both the original system $\Sys_1$ and the specifications $\Sigma_1$ are transposed into an \emph{abstract} domain, resulting in an abstract system $\Sys_2$ and corresponding abstract specifications $\Sigma_2$. We refer to the original system as the \emph{concrete} system as opposed to the abstract system. 
Next, an abstract controller $\Cont_2$ is synthesized to solve this abstract control problem $(\Sys_2,\Sigma_2)$. 
Finally, in the third step, called \emph{concretization} as opposed to abstraction, a controller $\Cont_1$ that enforces $\Sigma_1$ for $\Sys_1$ is derived from the abstract controller.

The effectiveness of this approach stems from replacing the concrete system, often characterized by an infinite number of states, with a finite state system. 
This enables the use of powerful control tools from abstract control synthesis~\citep{belta2017formal,kupferman2001model}, such as those derived from graph theory, including methods like Dijkstra or the A-star algorithm.

The correctness of this approach is ensured by relating the concrete system with its abstraction by a simulation relation.
The notion of \emph{alternating simulation relation} ($\ASR$)~\citep[Definition 4.19]{tabuada2009verification} is crucial; indeed, it is proved in the seminal work of~\citet{alur1998alternating} that $\ASR$ is a sufficient condition for a system $\Sys_2$ to correctly represent a system $\Sys_1$, such that any controller $\Cont_2$ for $\Sys_2$ can be concretized into a valid controller $\Cont_1$ for $\Sys_1$.

However, it soon appeared that $\ASR$ has weaknesses: 
while it formally characterizes abstractions allowing for abstraction-based controller design, the exact way to reconstruct the concrete controller may be computationally expensive.

An important landmark in the abstraction-based control literature is~\citet{reissig2016feedback}. There, the authors introduce the stronger \emph{feedback refinement relation} ($\FRR$)~\citep[Def. V.2]{reissig2016feedback}, designed to alleviate these limitations. 
They show that when $\FRR$ holds between the abstract and concrete systems, one can simply concretize the controller by seamlessly plugging the abstract controller with a quantizer, see Figure~\ref{fig:concretization_augmented} (left).
While this leads to a straightforward concretization procedure, where the concrete controller merely reproduces the control input of the abstract controller, it also limits the class of implementable controllers to piecewise constant ones (see discussion in~\citet[Section 4.4]{calbert2024memoryless}).

Recently, several other simulation relations that refine the $\ASR$ have been proposed in the literature~\citep{liu2017robust, borri2018design, egidio2022state, calbert2024memoryless}, each with its own benefits. While these notions are well understood individually, their relationships and relative strengths are usually presented separately. As a result, the landscape of simulation relations for symbolic control has become increasingly fragmented, making it harder to compare or systematically select among them.

In this paper, we unify these existing relations by interpreting each of them through the specific \emph{control architecture} it induces. 
Motivated by this observation, we take a step back from the existing definitions of simulation relations and ask:
\begin{center}
    \emph{What is the connection between a particular
    simulation relation and the corresponding control architecture?}
\end{center}
We propose a unified framework that captures the different types of simulation relations found in the literature and allows for their comparison in terms of the control architectures they induce.
Our construction builds on the concept of an \emph{interface}, in the spirit of~\citet{girard2009hierarchical}, which stores the output of the abstract controller \(\Cont_2\), combines it with the observed state of the concrete system \(\Sys_1\), and produces a valid control input. This architecture is illustrated in Figure~\ref{fig:concretization_augmented} (right).
Consequently, the resulting concretization procedure exhibits a \emph{plug-and-play} structure, enabling the abstract controller to be connected to the original system via an interface, irrespective of the particular specification we seek to enforce on the original system.

Our main theoretical contribution is a theorem (Theorem~\ref{th:augmented_system}, Corollary~\ref{cor:concrete_controller_augmented}) that explicitly links simulation relations to the control architecture that the concrete system inherits from the abstract one, as a function of the simulation relation that links them. Thus, the simulation relation becomes one of the elements in the control engineer’s toolbox, with the
choice of the precise relation to be used depending on the control architecture that (s)he wants to or is able to implement in practice. 

{In contrast to previous works, which typically prove only that a relation implies a correct interface, we also establish the converse: if the interface works for \emph{any} abstract controller, then the corresponding simulation relation must hold. To our knowledge, this two-way characterization was previously established only for the $\FRR$~\citep[Theorems V.4, V.5]{reissig2016feedback}. We leverage this result by proving that the \emph{interfaced system}—that is, the system composed of the concrete system and the interface (see Figure~\ref{fig:concretization_augmented}, right)—is always in feedback refinement relation with the abstract system.

Finally, as a by-product, we introduce new simulation relations that we believe are useful in practice, and we provide their explicit constructions on a common example, highlighting their respective advantages and drawbacks in terms of concretization procedures.

\begin{figure}
    \centering
    \includegraphics[width=0.48\textwidth]{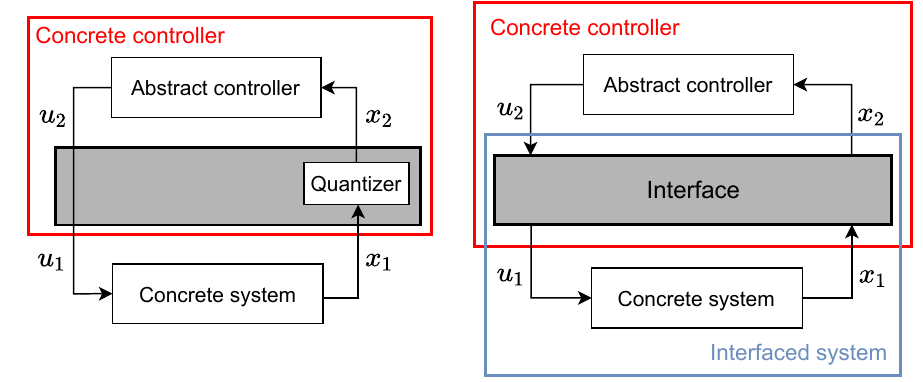}
    \caption{
    Closed-loop system resulting from the abstraction and concretization approach.
    Left: Based on the feedback refinement relation.
    Right: Based on the generic hierarchical control architecture used in this paper.}
    \label{fig:concretization_augmented}%
\end{figure}

\noindent \textbf{Notation}:
The sets $\R,\Z, \N$ denote respectively the sets of real numbers, integers and natural numbers.
For example, we use $[a,b]\subseteq \R$ to denote a closed continuous interval and $[a;b] = [a,b]\cap \Z$ for discrete intervals. The symbol $\myemptyset$ denotes the empty set.
Given two sets $A,B$, we define a \emph{single-valued map} as $f:A\rightarrow B$, while a~\emph{set-valued map} is defined as $f:A\rightarrow 2^B$, where $2^B$ is the power set of $B$, i.e., the set of all subsets of $B$. 
We denote the identity map $A\rightarrow A:a\rightarrow a$ by $I_A$.
The image of a subset $\Omega\subseteq A$ under $f:A\rightarrow 2^B$ is denoted $f(\Omega)=\bigcup_{a\in\Omega} f(a)$.
%
The set of maps $A\rightarrow B$ is denoted $B^A$, and the set of all signals that take their values in $B$ and are defined on intervals of the form $\length{N}=[0;N-1]$ is denoted $B^\infty$, $B^\infty = \bigcup_{N\in \N\cup \{\infty\}}B^{\length{N}}$. 
Given a set-valued map $f:X\rightarrow 2^{Y}$ and $\seq{x}\in X^{\length{N}}$, we denote by 
$f(\seq{x}) = \{\seq{y}\in Y^{\length{N}} \mid \forall k\in \length{N}:\ y(k)\in f(x(k))\}$.
Let $X\subseteq (V\times W)^\infty$, we define the projection onto the first component of the sequences as
\begin{multline}\label{eq:projection}
    \pi_{V}(X) = \{\seq{v}\in V^{\length{N}} \mid \exists \seq{w}\in W^{\length{N}}, \text{ such that }\\
    ((v(0),w(0)),\ldots,(v(N-1),w(N-1)))\in X\\
    \text{ for }N\in \N\cup\{\infty\}\}.
\end{multline}
We define $\pi_{W}(X)$ similarly.


\section{Control framework}\label{sec:control_framework}

In this section, we introduce the control framework used in this paper, as proposed in~\citet[Definition III.1]{reissig2016feedback}, which provides a unified formalism to describe systems, controllers, quantizers, and interconnected systems.

\begin{definition}\label{def:sys_bis}
    A system is a sextuple 
    \begin{equation}\label{eq:sys}
    \Sys = (\set{X},\set{U}, \set{V},\set{Y},F,H),
    \end{equation}
    where $\set{X}$, $\set{U}$, $\set{V}$, $\set{Y}$ are respectively the sets of \emph{states}, \emph{inputs}, \emph{internal variables} and \emph{outputs}, and $F:\set{X}\times \set{V}\rightarrow 2^{\set{X}}$ and $H:\set{X}\times \set{U}\rightarrow 2^{\set{Y}\times \set{V}}$ are respectively the \emph{transition} and \emph{output} maps
    such that
    \begin{align}
        x(k+1) &\in F(x(k), v(k)) \label{eq:sys_1}\\
        (y(k),v(k)) & \in H(x(k), u(k)).\label{eq:sys_2}
    \end{align}
    A quadruple $(\seq{u},\seq{v},\seq{x},\seq{y})\in \set{U}^{\length{N}} \times \set{V}^{\length{N}}\times \set{X}^{\length{N}} \times \set{Y}^{\length{N}}$ is a \emph{trajectory} of length $N$ of system $\Sys$ starting at $x(0)\in\set{X}$ if $N\in \N \cup\{\infty\}$, \eqref{eq:sys_1} holds for all $k\in \length{N-1}$, \eqref{eq:sys_2} holds for all $k\in \length{N}$.
    \hfill $\triangle$
\end{definition}
The internal variable \(v\) is required to describe the interconnection of systems through set-valued output maps without introducing new state variables (see the discussion after~\citet[Definition III.1]{reissig2016feedback}).
While this additional structure is not required to describe the systems we aim to control, it is essential for representing the specific class of controllers with set-valued output maps, introduced in this paper, as it will be formalized in Proposition~\ref{def:concretized_controller_formalism}.

The use of a set-valued function to describe the transition and output maps of a system allows to model perturbations and any other kind of non-determinism in a unified formalism.

We define the \emph{behavior} of a system as its set of \emph{maximal} trajectories.
\begin{definition}
Given the system $\Sys$ in~\eqref{eq:sys}, the \emph{behavior} of $\Sys$ is the set $\set{B}(\Sys) = \{\seq{y}\mid \exists \seq{u},\seq{v},\seq{x}\text{ such that }(\seq{u},\seq{v},\seq{x},\seq{y})$  is a trajectory of length $N$ of $\Sys$ on $[0;N[$ and if $N<\infty, \text{ then } F(x(N-1),v(N-1)) = \myemptyset \}$.
\hfill $\triangle$
\end{definition}

\begin{figure}
    \centering
    \begin{subfigure}{0.40\textwidth}
        \centering
        \includegraphics[width=\textwidth, trim=5 0 5 0, clip]{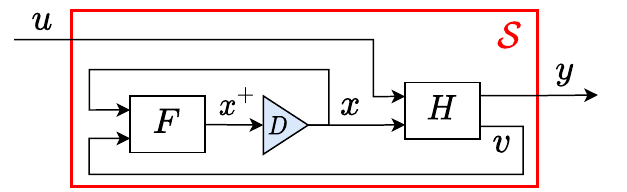}
        \caption{General system~\eqref{eq:sys}.}
        \label{fig:general_system}
    \end{subfigure}\\
    \begin{subfigure}{0.20\textwidth}
        \centering
        \includegraphics[width=\textwidth, trim=18 20 42 5, clip]{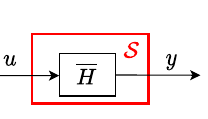}
        \caption{Static system~\eqref{eq:static_system}.}
        \label{fig:static_system}
    \end{subfigure}
    \hfill
    \begin{subfigure}{0.27\textwidth}
        \centering
        \includegraphics[width=\textwidth, trim=10 0 14 0, clip]{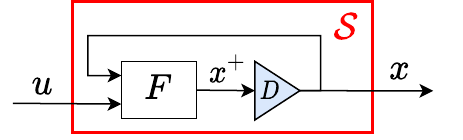}
        \caption{Simple system~\eqref{eq:simple_system}.}
        \label{fig:simple_system}
    \end{subfigure}
    \caption{System representations. Observe that (b) and (c) are particular cases of (a). The blue blocks are \emph{delay blocks} which represent memory elements that store and provide the previous value of a signal, e.g., $D(x(k)) = x(k-1)$.}
    \label{fig:systems}
\end{figure}
We classify the system~\eqref{eq:sys} as follows.
\begin{definition}
The system $\Sys$ in~\eqref{eq:sys} is 
\begin{enumerate}
    \item[$(i)$] \emph{Autonomous} if \(\set{U}\) is a singleton\footnote{When \(\set{X}\), \(\set{U}\), \(\set{V}\), and \(\set{Y}\) are singleton sets, we will consider, without loss of generality, the singleton \(\{0\}\).};
    \item[(ii)] \emph{Static} if \(\set{X}\) is a singleton, i.e.,
    \begin{equation}\label{eq:static_system}
        \Sys = (\{0\}, \set{U}, \set{V}, \set{Y}, F, H),
    \end{equation}
    with \(F(0, v) = \{0\}\) for all $v\in\set{V}$.
    \item[$(iii)$]
    \emph{Simple} if \(\set{V} = \set{U}\), \(\set{Y} = \set{X}\), and \(H = I_{\set{X}\times \set{U}}\), i.e.,
    \begin{equation}\label{eq:simple_system}
    \Sys = (\set{X}, \set{U}, \set{U}, \set{X}, F, I_{\set{X}\times \set{U}}).
    \end{equation}
    \end{enumerate}
\hfill $\triangle$
\end{definition}

 \begin{rem}
Intuitively, a static system captures an input-output behavior governed by
\begin{equation}\label{eq:static_sys}
    y(k) \in \overline{H}(u(k)),
\end{equation}
where the set-valued map \(\overline{H} : \set{U} \rightarrow 2^{\set{Y}}\) is defined as \(\overline{H}(u) = \{y \mid \exists v:\ (y, v) \in H(0, u)\}\). For the sake of simplifying notation, we denote the static system~\eqref{eq:static_sys} by \(\overline{H}\).
In contrast, a simple system represents a fully state-observable system with dynamics
\begin{equation}\label{eq:sys_no_output}
    x(k+1) \in F(x(k), u(k)).
\end{equation}
For the sake of simplifying notation, we denote the simple system~\eqref{eq:sys_no_output} by the tuple \(\Sys = (\set{X}, \set{U}, F)\).
We illustrate systems~\eqref{eq:sys}, \eqref{eq:static_system} and \eqref{eq:simple_system} in Figure~\ref{fig:systems}.
\end{rem}

For a simple system \(\Sys=(\set{X},\set{U},F)\), we introduce the set-valued operator of \emph{available inputs}, defined as \(\set{U}_{\mathcal{\Sys}}(x) = \{ \vect u \in \set{U} \mid F(x, u) \ne \varnothing \}\), which gives the set of inputs \(u\) available at a given state \(x\).
We say that a simple system is \emph{deterministic} if for every state \(x \in \set{X}\) and control input \(u \in \set{U}\), \(F(x, u)\) is either empty or a singleton. Otherwise, we say that it is \emph{non-deterministic}.

Given any two sets \(\set{U}\) and \(\set{Y}\), we identify a binary \emph{relation} \(R \subseteq \set{U} \times \set{Y}\) with set-valued maps, i.e., \(R(u) = \{y \mid (u, y) \in R\}\) and \(R^{-1}(y) = \{u \mid (u, y) \in R\}\).
Depending on the context, we will use the symbol \(R\) interchangeably to denote the binary relation itself, the associated set-valued map \(R : \set{U} \rightarrow 2^{\set{Y}}\), and its associated static system~\eqref{eq:static_system}, which is then referred to as a quantizer.


\begin{figure}
    \centering
    \includegraphics[width=0.35\textwidth]{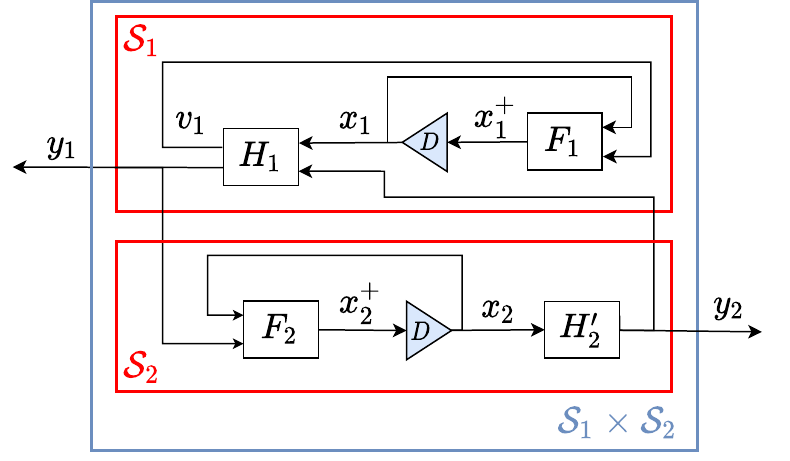}
    \caption{Feedback composition $\Sys_1\times \Sys_2$ of systems $\Sys_1$ and $\Sys_2$ in~\eqref{eq:sys} where $H_2'(x) = \{y_2\mid \exists u_2\in\set{U}_2:\ (y_2,u_2)\in H_2(x_2,u_2)\}$ (see Definition~\ref{def:feedback_composition}).
    }
    \label{fig:system_compositions}%
\end{figure}
We consider the \emph{serial} and the \emph{feedback composition} of two systems.
We start by defining the serial composition of two systems~ \citep[Definition III.2]{reissig2016feedback}.
\begin{definition}\label{def:serial_composition}
    Let $\Sys_i= (\set{X}_i,\set{U}_i, \set{V}_i,\set{Y}_i,F_i,H_i)$
    be systems, $i\in\{1,2\}$.
    Then, \(\Sys_1\) is \emph{serial composable} with \(\Sys_2\) if \(\set{Y}_1 \subseteq \set{U}_2\). The \emph{serial composition} of \(\Sys_1\) and \(\Sys_2\), denoted by \(\Sys_2 \circ \Sys_1\), is defined as
    \begin{equation}\label{eq:serial_composition}
        \Sys_2 \circ \Sys_1 = (\set{X}_{12}, \set{U}_1, \set{V}_{12}, \set{Y}_2, F_{12}, H_{12}),
    \end{equation}
    where \(\set{X}_{12} = \set{X}_1 \times \set{X}_2\), \(\set{V}_{12} = \set{V}_1 \times \set{V}_2\), \(F_{12} : \set{X}_{12} \times \set{V}_{12} \rightarrow 2^{\set{X}_{12}}\), and \(H_{12} : \set{X}_{12} \times \set{U}_1 \rightarrow 2^{\set{Y}_2\times \set{V}_{12}}\) satisfy
    \begin{align*}
        F_{12}(\vect{x}, \vect{v}) &= F_1(x_1, v_1) \times F_2(x_2, v_2),\\
        H_{12}(\vect{x}, u_1) &= \{(y_2, \vect{v}) \mid \exists y_1:\  (y_1, v_1) \in H_1(x_1, u_1) \\
        &\hspace{0.65cm}\land (y_2, v_2) \in H_2(x_2, y_1)\}.
    \end{align*}
\end{definition}
We define the feedback composition of two systems~\citep[Definition III.3]{reissig2016feedback} as illustrated in Figure~\ref{fig:system_compositions}.
\begin{definition}\label{def:feedback_composition}
Let \(\Sys_i = (\set{X}_i, \set{U}_i, \set{V}_i, \set{Y}_i, F_i, H_i)\) for \(i \in \{1,2\}\). Then, \(\Sys_1\) is \emph{feedback composable} with \(\Sys_2\), denoted as \(\Sys_1\) is f.c. with \(\Sys_2\), if \(\set{Y}_2 \subseteq \set{U}_1\), \(\set{Y}_1 \subseteq \set{U}_2\), and the following hold:  
(i) \(\set{V}_2 = \set{U}_2\) and \( H_2(x_2, u_2) = H_2'(x_2) \times \{u_2\} \) for some \( H_2': \set{X}_2 \rightarrow 2^{\set{Y}_2} \);  
(ii) If \((y_1, v_1) \in H_1(x_1, y_2)\), \((y_2, v_2) \in H_2(x_2, y_1)\), and \( F_2(x_2, v_2) = \myemptyset \), then \( F_1(x_1, v_1) = \myemptyset \).
The \emph{feedback composition}, denoted by \(\Sys_1 \times \Sys_2\), is defined as
    \begin{equation}\label{eq:feedback_composition}
        \Sys_1 \times \Sys_2 = (\set{X}_{12}, \{0\}, \set{V}_{12}, \set{Y}_{12}, F_{12}, H_{12}),
    \end{equation}
    where \(\set{X}_{12} = \set{X}_1 \times \set{X}_2\), \(\set{V}_{12} = \set{V}_1 \times \set{V}_2\), $\set{Y}_{12} = \set{Y}_1\times \set{Y}_2$, \(F_{12} : \set{X}_{12} \times \set{V}_{12} \rightarrow 2^{\set{X}_{12}}\), and \(H_{12} : \set{X}_{12} \times \{0\} \rightarrow 2^{\set{Y}_{12}\times \set{V}_{12}}\) satisfy
    \begin{align*}
        F_{12}(\vect{x}, \vect{v}) &= F_1(x_1, v_1) \times F_2(x_2, v_2),\\
        H_{12}(\vect{x}, 0) &= \{(\vect{y}, \vect{v}) \mid (y_1, v_1) \in H_1(x_1, y_2) \\
        &\hspace{0.65cm}\land (y_2, v_2) \in H_2(x_2, y_1)\}.
    \end{align*}
\end{definition}
Condition (i) in Definition~\ref{def:feedback_composition} prevents delay-free cycles~\citep{vidyasagar1981input} in the closed-loop system by introducing a one-step delay. Condition (ii) will be needed later to ensure that if the concrete closed loop is non-blocking, then so is the abstract closed loop.

In our context, the system $\Sys_2$ represents the system we aim to control, while $\Sys_1$ represents the controller.
Note that the system $\Sys_1\times \Sys_2$ is autonomous.

Given $\Sys_1$ and $\Sys_2$ in Definition~\ref{def:feedback_composition}, we have $\set{B}(\Sys_1 \times \Sys_2) \subseteq (\set{Y}_1 \times \set{Y}_2)^{\infty}$. To simplify further developments, we define the set $\B(\Sys_1 \times \Sys_2)$, which contains the output trajectories of $\Sys_2$ under feedback composition with $\Sys_1$, as follows
\begin{equation}
    \B(\Sys_1 \times \Sys_2) = \pi_{\set{Y}_2}(\set{B}(\Sys_1 \times \Sys_2)),
\end{equation}
where $\pi_{\set{Y}_2}(\cdot)$ is defined in~\eqref{eq:projection}.

We now define the system \emph{specifications} and, since we are focused on controller synthesis, the concept of a \emph{control problem}.
\begin{definition}\label{def:specification}
Consider the system $\Sys$ in~\eqref{eq:sys}. A \emph{specification} $\Sigma$ for $\Sys$ is defined as any subset $\Sigma \subseteq \set{Y}^{\infty}$. 
A system $\Sys$ together with a specification $\Sigma$ constitute a \emph{control problem} $(\Sys,\Sigma)$.
Additionally, a system $\Cont$, called the \emph{controller}, is said to \emph{solve} the control problem $(\Sys,\Sigma)$ if $\Cont$ is f.c. with $\Sys$ and $\B(\Cont\times \Sys)\subseteq \Sigma$.
    \hfill $\triangle$
\end{definition}
We can use linear temporal logic~\citep[Chapter 5]{baier2008principles} to define a specification $\Sigma$ for a system~$\Sys$.

\section{Characterization of simulation relations}\label{sec:prob_formulation}

\subsection{Abstraction-based control}

We now formalize the three-step abstraction-based control procedure described in the introduction (see Figure~\ref{fig:abstraction-based-control}) within our mathematical framework.

This work focuses on \emph{state-feedback} control using an abstraction-based approach. For this reason, both the system to be controlled and its abstraction are described as simple systems, while controllers require the more general formalism of Definition~\ref{def:sys_bis}, as it will be stated in Proposition~\ref{def:concretized_controller_formalism}. Specifically, we consider
\begin{align}
\Sys_i &=(\set{X}_i, \set{U}_i,F_i), \label{eq:simple_sys}\\
\Cont_i &=(\set{X}_{\Cont_i}, \set{X}_i,\set{V}_{\Cont_i}, \set{U}_i,F_{\Cont_i},H_{\Cont_i}),\label{eq:cont}
\end{align}
where $\Sys_1$ is the \emph{concrete system} and $\Sys_2$ its \emph{abstraction}.

%
To solve the concrete control problem $(\Sys_1, \Sigma_1)$ with $\Sigma_1\subseteq \set{X}_1^{\infty}$, the abstraction-based control procedure proceeds as follows.
In step 1, the system $\Sys_1$ is abstracted into a system $\Sys_2$, meaning their states are related through a relation $R \subseteq \set{X}_1 \times \set{X}_2$. Additionally, the concrete specification is abstracted into an abstract specification $\Sigma_2\subseteq \set{X}_2^{\infty}$, satisfying
\begin{equation}\label{eq:abstract_spec}
R^{-1}(\Sigma_2)\subseteq \Sigma_1.
\end{equation}
%
Step 2 entails designing a controller $\Cont_2$ that solves the abstract problem $(\Sys_2,\Sigma_2)$, i.e., such that
\begin{equation}\label{eq:abstract_problem}
\B(\Cont_2\times \Sys_2) \subseteq\Sigma_2.
\end{equation}
Finally, step 3 involves constructing a controller $\Cont_1$ that satisfies the \emph{reproducibility} as defined in~\citet[Property 1]{calbert2024memoryless} 
\begin{equation}\label{eq:third_step}
\B(\Cont_1 \times \Sys_1)\subseteq  R^{-1}(\B(\Cont_2 \times \Sys_2)).
\end{equation}
The validity of this approach is established by the following inclusions, which guarantee that the controller $\Cont_1$ solves the concrete problem $(\Sys_1, \Sigma_1)$
$$\B(\Cont_1 \times \Sys_1)\subseteq  R^{-1}(\B(\Cont_2 \times \Sys_2))\subseteq R^{-1}(\Sigma_2)\subseteq \Sigma_1$$
which follows directly from~\eqref{eq:third_step},~\eqref{eq:abstract_problem},~\eqref{eq:abstract_spec}, and the monotonicity of $R^{-1}$.


In order to be able to construct $\Cont_1$ from $\Cont_2$  satisfying~\eqref{eq:third_step}, the relation $R$ must impose conditions on the local dynamics of the systems in the associated states, accounting for the impact of different inputs on state transitions. 
The \emph{alternating simulation relation}~\citep[Definition 4.19]{tabuada2009verification} is a comprehensive definition of such a relation. 
\begin{definition}\label{def:ASR}
Given two simple systems $\Sys_1$ and $\Sys_2$ in~\eqref{eq:simple_sys}, a relation $R \subseteq \set{X}_1\times \set{X}_2$ is an \emph{alternating simulation relation} from $\Sys_1$ to $\Sys_2$, denoted $\Sys_1\preceq_R^{\ASR} \Sys_2$, if
\vspace{-0.9cm}
{\small \begin{multline}\label{eq:ASR_def}
\forall (x_{1},x_{2}) \in R \ \forall u_2\in \set{U}_{\Sys_2}(x_2)\ \exists u_1\in \set{U}_{\Sys_1}(x_1)\\  
    \forall x_1^+\in F_1(x_1,u_1)\  \exists x_2^+\in  F_2(x_2,u_2):\ (x_1^+,x_2^+)\in R.
\end{multline}}
\end{definition}
The alternating simulation relation guarantees that any controller applied to $\Sys_2$ can be concretized into a controller for $\Sys_1$ that satisfies~\eqref{eq:third_step}.
\begin{theorem}[{\citet[Proposition 8.7]{tabuada2009verification}}]\label{th:ASR_concretization_property}
Given two simple systems $\Sys_1$ and $\Sys_2$ as in~\eqref{eq:simple_sys}, if $\Sys_1 \preceq_R^{\ASR} \Sys_2$, then for every controller $\Cont_2$ f.c. with $\Sys_2$, there exists a controller $\Cont_1^{\ASR}$ satisfying $\B(\Cont_1^{\ASR} \times \Sys_1)\subseteq  R^{-1}(\B(\Cont_2 \times \Sys_2))$.
\end{theorem}

Although the alternating simulation relation provides a safety-critical framework, it has several practical drawbacks in the implementation of the concrete controller $\Cont_1$, which we will discuss in Section~\ref{sec:discussion}. To address these issues, various alternative relations have been introduced in the literature. Among these, the \emph{feedback refinement relation}~\citep[Definition V.2]{reissig2016feedback} is central to the further developments of this paper.

\begin{definition}\label{def:FRR}
Given two simple systems $\Sys_1$ and $\Sys_2$ as in~\eqref{eq:simple_sys}, a relation $R \subseteq \set{X}_1\times \set{X}_2$ is a \emph{feedback refinement relation} from $\Sys_1$ to $\Sys_2$, denoted $\Sys_1\preceq_R^{\FRR} \Sys_2$, if 
 for every $(x_{1},x_{2}) \in~R$
 \begin{align}
    (i)\ &\set{U}_{\Sys_2}(x_2)\subseteq \set{U}_{\Sys_1}(x_1);\nonumber\\
    (ii)\ &\forall u_2\in\set{U}_{\Sys_2}(x_2)\ 
        \forall x_1^+\in F_1(x_1, u_2): \ R(x_1^+)~\ne~\myemptyset\nonumber\\ 
    &\text{and } \ \forall x_2^+\in R(x_1^+):\  x_2^+\in F_2(x_2, u_2).\label{eq:FRR_def}
 \end{align}
\end{definition}
As a refined notion of $\ASR$, the feedback refinement relation benefits from a straightforward concretization scheme as established by the following theorem.
\begin{theorem}[{\footnotesize \citet[Thms V.4, V.5]{reissig2016feedback}}]
\label{th:FRR_concretization}
Consider the simple systems $\Sys_1$ and $\Sys_2$ as in~\eqref{eq:simple_sys}. Given $R \subseteq \set{X}_1 \times \set{X}_2$, the following equivalence holds:
\begin{enumerate}
    \item $\Sys_1 \preceq_{R}^{\FRR} \Sys_2$;
    \item For all $\Cont_2$ f.c. with $\Sys_2$: $\Cont_2$ is f.c. with $R \circ \Sys_1$, $\Cont_2 \circ R$ is f.c. with $\Sys_1$, and $\B(\Cont_2 \times (R \circ \Sys_1)) \subseteq \B(\Cont_2 \times \Sys_2)$.
\end{enumerate}
Additionally, it implies that
\begin{equation}\label{eq:concrete_controller_FRR_def}
    \Cont_1^{\FRR} \defEqual \Cont_2 \circ R,
\end{equation}
satisfies
$\B(\Cont_1^{\FRR} \times \Sys_1) \subseteq R^{-1}(\B(\Cont_2 \times \Sys_2))$.
\end{theorem}



The concretization procedure for \(\FRR\), defined as the serial composition of the abstract controller \(\Cont_2\) with the quantizer \(R\)~\eqref{eq:concrete_controller_FRR_def}, exhibits a plug-and-play property: the control architecture (Figure~\ref{fig:concretization_augmented} (left)) remains independent of the specification to be enforced on the original system. However, this advantage comes at a cost—it constrains how the abstraction is constructed, as it forces the concretized controller to be piecewise constant. In particular, conditions (i) and (ii) in Definition~\ref{def:FRR} restrict the class of concrete controllers to those where the control input depends solely on the abstract state~\eqref{eq:concrete_controller_FRR_def}.

Thus, the feedback refinement relation is well-suited to contexts where the exact state is unknown and only abstract (or symbolic) state information is available. However, when information on the concrete state is available, this becomes a significant limitation.
If the system does not exhibit local incremental stability~\citep[Definition 2.1]{angeli2002lyapunov}\citep{lohmiller1998contraction}, meaning that trajectories move away from each other, the use of piecewise constant controllers introduces significant non-determinism into the abstraction. This could result in an intractable or even unsolvable abstract problem \((\Sys_2, \Sigma_2)\), as illustrated by the example in~\citet[Section 4.4]{calbert2024memoryless}.

Therefore, our goal is to identify, classify, and characterize refined notions of the alternating simulation relation that allow the design of more general concrete controllers than the piecewise constant controllers of the feedback refinement relation, while still benefiting from its plug-and-play property.

\subsection{Key relations}\label{sec:simulation_relations}
In this section, we introduce refined notions of alternating simulation relations, which will be discussed throughout this paper. The first two relations are introduced in this paper, while the last is presented and discussed in~\citet[Definition 4]{liu2017robust} and \citet[Definition 8]{calbert2024memoryless}.

\begin{definition}\label{def:ASRB}
Given two simple systems $\Sys_1$ and $\Sys_2$ in~\eqref{eq:simple_sys}, a relation $R \subseteq \set{X}_1\times \set{X}_2$ is a \emph{predictive simulation relation} from $\Sys_1$ to $\Sys_2$, denoted $\Sys_1\preceq_R^{\ASRB} \Sys_2$, if 
\vspace{-0.9cm}
{\small \begin{multline}\label{eq:ASRB_def}
    \forall (x_1,x_2)\in R \ \forall u_2\in \set{U}_{\Sys_2}(x_2)\ \exists u_1\in\set{U}_{\Sys_1}(x_1)\\ \exists x_2^+\in F_2(x_2, u_2) \
    \forall x_1^+\in F_1(x_1, u_1): \ (x_1^+,x_2^+)\in R.
\end{multline}}
\end{definition}

\begin{definition}\label{def:ASRBB}
Given two simple systems $\Sys_1$ and $\Sys_2$ in~\eqref{eq:simple_sys}, 
a relation $R \subseteq \set{X}_1\times \set{X}_2$ is a \emph{feedforward abstraction relation} from $\Sys_1$ to $\Sys_2$, denoted $\Sys_1\preceq_R^{\ASRBB} \Sys_2$, if 
\vspace{-0.9cm}
{\small \begin{multline}\label{eq:ASRBB_def}
    \forall x_2\in \set{X}_2\ \forall u_2\in \set{U}_{\Sys_2}(x_2)\ \exists x_2^+ \in F_2(x_2,u_2)\ 
    \forall x_1\in R^{-1}(x_2)\\
    \exists u_1\in \set{U}_{\Sys_1}(x_1)\ \forall x_1^+\in F_1(x_1, u_1):\ (x_1^+,x_2^+)\in R.
\end{multline}}
\end{definition}

\begin{definition}\label{def:MCR}
Given two simple systems $\Sys_1$ and $\Sys_2$ in~\eqref{eq:simple_sys}, a relation $R \subseteq \set{X}_1\times \set{X}_2$ is a \emph{memoryless concretization relation} from $\Sys_1$ to $\Sys_2$, denoted $\Sys_1\preceq_R^{\MCR} \Sys_2$,~if $\forall (x_1,x_2)\in R$
\vspace{-0.9cm}
{\small \begin{multline}\label{eq:MCR_def}
    \forall u_2\in \set{U}_{\Sys_2}(x_2)\ \exists u_1\in\set{U}_{\Sys_1}(x_1)\
       \forall x_1^+\in F_1(x_1, u_1)\\ \forall x_2^+\in R(x_1^+):\ x_2^+\in F_2(x_2, u_2).
\end{multline}}
\end{definition}

We define the associated extended relations $R^{\ASR}_e$, $R^{\ASRB}_e$, $R^{\MCR}_e \subseteq \set{X}_2\times \set{U}_2\times \set{X}_1\times \set{U}_1$ as the sets of tuples $(x_2, u_2, x_1, u_1)$ satisfying~\eqref{eq:ASR_def}, \eqref{eq:ASRB_def}, and \eqref{eq:MCR_def} respectively. Similarly, $R^{\ASRBB}_e\subseteq \set{X}_2\times \set{U}_2\times \set{X}_1\times \set{U}_1\times \set{X}_2$ is defined as the set of tuples $(x_2, u_2, x_1, u_1, x_2^+)$ satisfying~\eqref{eq:ASRBB_def}.
In addition, to extract the corresponding concrete inputs, we introduce the following set-valued maps
\begin{equation}\label{eq:interface_abstract_concrete_input}
I_R^{\T}(x_2, u_2, x_1) = \{u_1 \mid (x_2, u_2, x_1, u_1) \in R_e^{\T}\},
\end{equation}
for \(\T \in \{\ASR, \ASRB, \MCR\}\), $I_R^{\ASRBB}(x_2, u_2, x_1, x_2^+) = \{u_1 \mid (x_2, u_2, x_1, u_1, x_2^+) \in R_e^{\ASRBB}\}$, $I_R^{\FRR}(u_2) = \{u_2\}$.

Intuitively, these maps ensure that each abstract input \(u_2\) can be matched with a concrete input \(u_1\) that satisfies the formula defining the relation.

\definecolor{input1}{gray}{0.9} 
\definecolor{input2}{gray}{0.8} 
\definecolor{input3}{gray}{0.7} 

\newcommand{\eqsizetree}{\large}

\begin{figure}
    \centering
    \resizebox{0.27\textwidth}{!}{
\begin{tikzpicture}[
    node distance=1.8cm,
    every path/.style={->, thick, line width=0.32mm}, 
    >={Latex[length=3mm, width=3mm]} 
    ]
    \node (1) [circle, draw, fill=input1] {\eqsizetree $\hyperref[def:ASR]{\ASR}$};
    \node (2) [circle, below of=1] {};
    \node (3) [circle, below of=2] {};
    \node (4) [circle, draw, fill=input3, left of=2] {\eqsizetree $\hyperref[def:ASRB]{\ASRB}$};
    \node (5) [circle, below of=4] {};
    \node (6) [circle, draw, fill=input3, left of=5] {\eqsizetree $\hyperref[def:ASRBB]{\ASRBB}$};
    \node (7) [circle, draw, fill=input3, right of=2] {\eqsizetree $\hyperref[def:MCR]{\MCR}$};
    \node (8) [circle, below of=7] {};
    \node (9) [circle, draw, fill=input3, right of=8] {\eqsizetree $\hyperref[def:FRR]{\FRR}$};
    
    \draw[->, thick, line width=0.5mm] (4) -- (1);
    \draw[->, thick, line width=0.5mm] (6) -- (4);
    \draw[->, thick, line width=0.5mm] (7) -- (1);
    \draw[->, thick, line width=0.5mm] (9) -- (7);
    
    \draw[->, red, thick, line width=0.5mm, bend right=30] (1) to node[pos=0.5, above, sloped] {\eqsizetree $A_1$} (6);
    \draw[->, red, thick, line width=0.5mm, bend left=28] (1) to node[pos=0.5, above, sloped] {\eqsizetree $A_2$} (7);
    \draw[->, red, thick, line width=0.5mm, bend left=28] (7) to node[pos=0.5, above, sloped] {\eqsizetree $A_3$} (9);
\end{tikzpicture}}
    \caption{Pre-order of simulation relations. For example, an arrow from $\ASRB$ to $\ASR$ indicates that $\Sys_1 \preceq_R^{\ASRB} \Sys_2$ implies $\Sys_1 \preceq_R^{\ASR} \Sys_2$. Red arrows represent implications that hold under specific assumptions labeled on the arrows: $A_1$ denotes that $\Sys_2$ is deterministic, $A_2$ that $R$ is deterministic, and $A_3$ that $u_1=u_2$ as in Proposition~\ref{prop:meta_relations}.}
    \label{fig:meta_relations}
\end{figure}

The following proposition provides a partial order over the simulation relations, corresponding to the tree structure shown in Figure~\ref{fig:meta_relations}.
\begin{proposition}\label{prop:meta_relations}
    Given the simple systems $\Sys_1$ and $\Sys_2$ as in~\eqref{eq:simple_sys}, and~$R\subseteq \set{X}_1\times \set{X}_2$, the following implications hold
    \begin{enumerate}
        \item[(1)] $\Sys_1\preceq_R^{\ASRBB} \Sys_2\Rightarrow\Sys_1\preceq_R^{\ASRB} \Sys_2\Rightarrow \Sys_1\preceq_R^{\ASR} \Sys_2$.
        \item[(2)] $\Sys_1\preceq_R^{\FRR} \Sys_2\Rightarrow\Sys_1\preceq_R^{\MCR} \Sys_2\Rightarrow \Sys_1\preceq_R^{\ASR} \Sys_2$.
    \end{enumerate}
    Additionally, 
    \begin{enumerate}
        \item[(3)] If $\Sys_2$ is deterministic,   $\Sys_1\preceq_R^{\ASR} \Sys_2 \Leftrightarrow \Sys_1\preceq_R^{\ASRBB} \Sys_2.$
        \item[(4)] If $R$ is deterministic,  $\Sys_1\preceq_R^{\ASR} \Sys_2 \Leftrightarrow \Sys_1\preceq_R^{\MCR} \Sys_2$.
        \item[(5)] If $\Sys_1\preceq_R^{\MCR}\Sys_2$ and $u_1\in I_R^{\MCR}(x_2,u_2,x_1)\Rightarrow u_1 = u_2$, then $\Sys_1\preceq_R^{\MCR}\Sys_2\Leftrightarrow \Sys_1\preceq_R^{\FRR}\Sys_2$.
    \end{enumerate}
\end{proposition}
\begin{pf}
    \begin{enumerate}[align=left]
        \item[(1), (2)] These follow from basic logical manipulations.
        \item[(3)] If $\Sys_2$ is deterministic, then for any $x_2 \in \set{X}_2$ and $u_2 \in \set{U}_{\Sys_2}(x_2)$: $|F_2(x_2, u_2)| = 1$. Thus, \eqref{eq:ASR_def} is equivalent to:
        $\forall (x_1, x_2) \in R\ \forall u_2 \in \set{U}_{\Sys_2}(x_2)\ x_2^+ \in F_2(x_2, u_2)\ \exists u_1 \in \set{U}_{\Sys_1}(x_1)\ \forall x_1^+ \in F_1(x_1, u_1): \ (x_1^+, x_2^+) \in R,$
        which in turn is equivalent to~\eqref{eq:ASRBB_def}.
    \item[(4)] If $R$ is deterministic, then for any $(x_1,x_2)\in R$: $|R(x_1)| = 1$. Therefore $R(x_1^+)\cap F_2(x_2,u_2)\neq \myemptyset$ if and only if $R(x_1^+)\ne \myemptyset,\ R(x_1^+)\subseteq F_2(x_2,u_2)$.
    \item[(5)] 
    When the abstract and concrete inputs in the extended relation are identical, \eqref{eq:MCR_def} simplifies to \eqref{eq:FRR_def}.
    \end{enumerate}
    \hfill\hfill\qed
\end{pf}

\subsection{Main results}\label{sec:main_results}
In this section, we present our main results, which provide an exact characterization of simulation relations in terms of their associated concretization procedures. To this end, we introduce the notion of an \emph{interface}, which connects the system \(\Sys_1\) to any controller \(\Cont_2\) functionally compatible with \(\Sys_2\), as illustrated in Figure~\ref{fig:concretization_augmented} (right).
\begin{definition}\label{def:interface}
Given two simple systems \(\Sys_1\) and \(\Sys_2\) as in~\eqref{eq:simple_sys}, an \emph{interface} from $\Sys_2$ to $\Sys_1$ is defined by the tuple $(\set{Z}_1,h_1,h_2,\widetilde{R})$, where $\set{Z}_1$ is a set, $h_1$ and $h_2$ are set-valued functions, and $\widetilde{R}\subseteq (\set{X}_1\times \set{Z}_1)\times \set{X}_2$.
    Consider the set of variables \(\nu = \{x_1, x_1^+, z_1, z_1^+, u_1, u_2\}\), where \(x_1, x_1^+ \in \set{X}_1\), \(z_1, z_1^+ \in \set{Z}_1\), \(u_1 \in \set{U}_1\), and \(u_2 \in \set{U}_2\).
    Let \(h_1\) (resp. \(h_2\)) be a set-valued function with argument \(\nu_1\) (resp. \(\nu_2\)), a subset of the set \(\nu\), returning a subset of \(\set{U}_1\) (resp. \(\set{Z}_1\)). The domain of \(h_1\) (resp. \(h_2\)) is denoted as \(\set{V}_1\) (resp. \(\set{V}_2\)), defined as the Cartesian product of the domains of the corresponding variables, i.e., \(h_1: \set{V}_1 \rightarrow 2^{\set{U}_1}\) and \(h_2: \set{V}_2 \rightarrow 2^{\set{Z}_1}\).
    The functions \(h_1\) and \(h_2\) ensure that the following holds. For all \(((x_1, z_1), x_2) \in \widetilde{R}\) and \(u_2 \in \set{U}_{\Sys_2}(x_2)\), it holds that any $u_1$, \(z_1^+\), \(x_1^+\) and $x_2^+$ which are solution of the following equations
        \begin{equation}\label{eq:interface}
        \begin{aligned}
             u_1 \in h_1(\nu_1)&, \ 
            z_1^+ \in h_2(\nu_2), \\
             x_1^+ \in F_1(x_1, u_1)&, \ 
            x_2^+ \in \widetilde{R}((x_1^+,z_1^+)),
        \end{aligned}
    \end{equation}
    satisfy $x_2^+ \in F_2(x_2, u_2)$.
    \hfill $\triangle$
\end{definition}
Intuitively, the variable $z_1$ serves as a state of the interface, allowing it to retain memory across transitions. Given $((x_1, z_1), x_2) \in \widetilde{R}$, the function \(h_1\) selects a concrete input \(u_1\) such that the resulting concrete successor \(x_1^+ \in F_1(x_1, u_1)\) remains related to an abstract successor \(x_2^+ \in F_2(x_2, u_2)\) via a new interface state \(z_1^+ \in h_2(\nu_2)\), ensuring \(((x_1^+, z_1^+), x_2^+) \in \widetilde{R}\).
In this way, the interface coordinates the evolution of the concrete system so that it stays related to the abstract system for any input \(u_2 \in \set{U}_{\Sys_2}(x_2)\), i.e., for any abstract controller \(\Cont_2\) f.c. with~\(\Sys_2\).

We now introduce the notion of an \emph{interfaced system}, which encapsulates the system \(\Sys_1\) together with the interface, as illustrated in Figure~\ref{fig:augmented_system_bis}.
\begin{definition}\label{def:augmented_system}
      Given the simple systems $\Sys_1$ and $\Sys_2$ in~\eqref{eq:simple_sys}, and an interface $(\set{Z}_1, h_1, h_2, \widetilde{R})$ from $\Sys_2$ to $\Sys_1$, we define the associated \emph{interfaced system} $\widetilde{\Sys}_1$ as the simple system $\widetilde{\Sys}_1 = (\widetilde{\set{X}}_1, \widetilde{\set{U}}_1, \widetilde{F}_1)$, where $\widetilde{\set{X}}_1 = \set{X}_1 \times \set{Z}_1$, $\widetilde{\set{U}}_1 = \set{U}_2$, and
      \begin{align}\label{eq:augmented_system}
        \widetilde{F}_1((x_1, z_1), u_2) = \{(x_1^+, z_1^+) \mid& u_1\in h_1(\nu_1),\ z_1^+ \in h_2(\nu_2),\nonumber\\
        &x_1^+ \in F_1(x_1, u_1)\}.
    \end{align}
        \hfill $\triangle$
\end{definition}

Next, we define the controller \(\Cont_1\), referred to as the \emph{concretized controller}, which results from connecting the controller \(\Cont_2\) to the interface, as illustrated in Figure~\ref{fig:augmented_system_bis}.
\begin{definition}\label{def:concretized_controller}
    Given two simple systems \(\Sys_1\) and \(\Sys_2\) as in~\eqref{eq:simple_sys}, an interface $(\set{Z}_1,h_1,h_2,\widetilde{R})$ from $\Sys_2$ to $\Sys_1$, and a controller $\Cont_2$ in~\eqref{eq:cont} f.c. with $\Sys_2$, a \emph{concretized controller} is any system $\Cont_1$ in~\eqref{eq:cont} that satisfies the following closed-loop condition: $\seq{x}_1 \in \B(\Cont_1 \times \Sys_1)$ if and only if there exists $\seq{u}_1,\seq{z}_1, \seq{x}_2, \seq{u}_2, \seq{x}_{\Cont_2}, \seq{v}_{\Cont_2}$ such that
    \begin{equation}\label{eq:interface_closed_loop_augmented}
            \begin{aligned}
                \tikzmark{start10}(u_2(k),v_{\Cont_2}(k))&\in H_{\Cont_2}(x_{\Cont_2}(k),x_2(k))\\
                x_{\Cont_2}(k+1)&\in F_{\Cont_2}(x_{\Cont_2}(k),v_{\Cont_2}(k))\tikzmark{end10}\\
                \noalign{\vspace{-0.3cm}}
                & \\
                 \tikzmark{start11} u_1(k) &\in h_1(\nu_1(k))\\
                 z_1(k+1) &\in h_2(\nu_2(k))\\
                 x_2(k+1) &\in \widetilde{R}((x_1(k+1),z_1(k+1))) \tikzmark{end11}\\
                \noalign{\vspace{-0.25cm}}
                & \\
                \tikzmark{start12}x_1(k+1) &\in F_1(x_1(k), u_1(k)).\tikzmark{end12}
            \end{aligned}
        \end{equation}
            \hfill $\triangle$
        \begin{tikzpicture}[remember picture,overlay]
          \coordinate (start) at ([xshift=-0.1em,yshift=2.0ex]pic cs:start10);
          \coordinate (end) at ([xshift=2.75em, yshift=-0.2em]pic cs:end10);
          \node[inner sep=2.3pt, draw=black, thick, line width=0.20mm, fit=(start) (end)] {};
        \end{tikzpicture}
         \begin{tikzpicture}[remember picture,overlay]
          \coordinate (start) at ([xshift=-3.92em,yshift=1.7ex]pic cs:start11);
          \coordinate (end) at ([xshift=0.2em, yshift=-0.3em]pic cs:end11);
          \node[inner sep=2.3pt, draw=black, thick, line width=0.20mm, fit=(start) (end)] {};
        \end{tikzpicture}
        \begin{tikzpicture}[remember picture,overlay]
          \coordinate (start) at ([xshift=-2.24em,yshift=1.8ex]pic cs:start12);
          \coordinate (end) at ([xshift=3.63em, yshift=-0.3em]pic cs:end12);
          \node[inner sep=2.3pt, draw=black, thick, line width=0.20mm, fit=(start) (end)] {};
        \end{tikzpicture}
\end{definition}
Equation~\eqref{eq:interface_closed_loop_augmented} describes the interaction between the dynamics of the abstract controller, the interface, and the concrete system, respectively.

\definecolor{myblue}{RGB}{108, 142, 191}
\begin{figure}
\centering
\resizebox{0.48\textwidth}{!}{
\begin{tikzpicture}[
  node distance=1.2cm and 2.5cm,
  box/.style={draw, minimum width=4.5cm, minimum height=2.3cm, align=center},
  smallbox/.style={draw, minimum width=4.8cm, minimum height=1.2cm, align=center},
  ->, >=Stealth
]

\node[smallbox, minimum width=3.2cm, fill=gray!20] (T1) {$\Sys_1 \preceq_R^{\T} \Sys_2$};

\node[smallbox, right=of T1] (T2) {
$\widetilde{\Sys}_1^{\T} \preceq_{\widetilde{R}}^{\FRR} \Sys_2$
\\
(Definition~\ref{def:augmented_system})
};

\node[smallbox, below=1.6cm of T2] (T3) {
{\small $\forall \Cont_2:$}\\
{\small 
    $\B(\Cont_2\times (\widetilde{R}\circ \widetilde{\Sys}_1))\subseteq$}\\
{\small $\B(\Cont_2\times \Sys_2)$
}};

\node[smallbox, below=of T3, fill=gray!20] (T4) {
{\small $(\set{Z}_1,h_1^{\T}, h_2^{\T},\widetilde{R})$ is an 
interface}\\
(Definition~\ref{def:interface})};

\node[smallbox, minimum width=3.2cm, fill=gray!60, left=of T4, below=1.4cm of T1] (T5) {
{\small $\forall \Cont_2:$}\\
{\small 
$\B(\Cont_1^{\T}\times \Sys_1)\subseteq$}\\
{\small $R^{-1}(\B(\Cont_2\times \Sys_2))$}\\ (Definition~\ref{def:concretized_controller})};

\node[draw, dashed, fit=(T2)(T3), inner sep=6pt] (GroupBox) {};

\node at ($(GroupBox.north)+(0,0.3)$) {\textbf{Interfaced system–$\FRR$ Trick}};

\draw[->] 
  ($(T1.east)+(0,0.30)$) 
  -- node[midway, above]{{\small $\widetilde{R}$ in~\eqref{eq:R_to_Rtilde}}} 
  ($(GroupBox.west |- T1.east)+(0,0.30)$);

\draw[<-] 
  ($(T1.east)-(0,0.30)$) 
  -- node[midway, below]{{\small $R$ in~\eqref{eq:Rtilde_to_R}}} 
  ($(GroupBox.west |- T1.east)-(0,0.30)$);

\node at ($(T1.east)!0.5!(GroupBox.west |- T1.east)$) { Theorem~\ref{th:augmented_system}};

  
\draw[<->] (T2) -- node[midway, right]{Theorem~\ref{th:FRR_concretization}} (T3);

\draw[<->] 
  ($(GroupBox.south)$) 
  -- node[midway, right]{Theorem~\ref{th:concrete_controller_augmented}} 
  (T4.north);


\draw[<-]  
  (T5.east) 
  -- 
  node[midway, above]{Theorem~\ref{th:concrete_controller_augmented}} 
  node[midway, below]{{\small $R$ in~\eqref{eq:Rtilde_to_R}}}
  ($(GroupBox.west |- T5.east)$);


\draw[->] 
  (T1) -- 
  node[midway, anchor=west, xshift=-6pt]{
    \begin{tabular}{c}
      Corollary~\ref{cor:concrete_controller_augmented} \\[-3pt]
      {\small $\widetilde{R}$ in~\eqref{eq:R_to_Rtilde}}
    \end{tabular}
  } 
  (T5);

\end{tikzpicture}
}
\caption{
Summary of the contributions in Section~\ref{sec:main_results}. The diagram illustrates how the main theorems and definitions are connected to characterize a simulation relation \(\T\) in terms of a specific control architecture, defined by the interface \((\set{Z}_1, h_1^{\T}, h_2^{\T}, \widetilde{R})\). It also shows how the resulting concretized controller \(\Cont_1^{\T}\) ensures the reproducibility condition~\eqref{eq:third_step}. 
}
\label{fig:lift}
\end{figure}

In order to establish the equivalence between a simulation relation and an interface, we will show that the interfaced system is always in a feedback refinement relation with the abstract one, as illustrated in Figure~\ref{fig:augmented_system_bis}. This insight will allow us to leverage the two-way characterization of \(\FRR\) (Theorem~\ref{th:FRR_concretization}) and form the core of our proof strategy, summarized in Figure~\ref{fig:lift}.

The following theorem shows that an interface induces a feedback refinement relation between the interfaced system and the abstract one, and formalizes the role of interfaces in the controller concretization process.
\begin{theorem}\label{th:concrete_controller_augmented}
    Consider two simple systems $\Sys_1$ and $\Sys_2$ as in~\eqref{eq:simple_sys}, and the interfaced system $\widetilde{\Sys}_1$ (Definition \ref{def:augmented_system}) associated with the tuple $(\set{Z}_1,h_1,h_2,\widetilde{R})$.
    The following propositions are equivalent.
\begin{enumerate}
    \item[(1)] $(\set{Z}_1,h_1,h_2,\widetilde{R})$ is an interface from $\Sys_2$ to $\Sys_1$;
    \item[(2)] $\widetilde{\Sys}_1\preceq_{\widetilde{R}}^{\FRR}\Sys_2$;
    \item[(3)] for any controller $\Cont_2$ f.c. with $\Sys_2$, it holds that $\Cont_2$ is f.c. with $\widetilde{R}\circ \widetilde{\Sys}_1$ and $\widetilde{\Cont_1}=\Cont_2\circ \widetilde{R}$ is f.c. with $\widetilde{\Sys}_1$, and  $\B(\Cont_2 \times (\widetilde{R} \circ \widetilde{\Sys}_1)) \subseteq \B(\Cont_2 \times \Sys_2)$.
\end{enumerate}
Additionally, for any controllers $\Cont_2$ f.c. with $\Sys_2$, the concretized controller $\Cont_1$ (Definition \ref{def:concretized_controller}) satisfies~\eqref{eq:third_step}, i.e., $\B(\Cont_1 \times \Sys_1) \subseteq R^{-1}(\B(\Cont_2 \times \Sys_2))$, with 
\begin{equation}\label{eq:Rtilde_to_R}
    R = \{(x_1,x_2)\mid \exists z_1\in \set{Z}_1 \text{ s.t. }((x_1,z_1), x_2)\in \widetilde{R}\}.
    \end{equation}
\end{theorem}
\begin{pf}
We first establish the equivalences. \\
    $\bf{(1)\Leftrightarrow (2)}$: The tuple \((\set{Z}_1, h_1, h_2, \widetilde{R})\) is an interface if and only if it satisfies~\eqref{eq:interface}. This is equivalent to requiring that for any \(((x_1, z_1), x_2) \in \widetilde{R}\) and \(u_2 \in \set{U}_{\Sys_2}(x_2)\), $(i)$ \(\widetilde{F}_1((x_1, z_1), u_2) \ne \myemptyset\), and $(ii)$ for all \((x_1^+, z_1^+) \in \widetilde{F}_1((x_1, z_1), u_2)\), \(\widetilde{R}((x_1^+, z_1^+)) \ne \myemptyset\) and \(\widetilde{R}((x_1^+, z_1^+)) \subseteq F_2(x_2, u_2)\). This, in turn, is equivalent to conditions $(i)$ and $(ii)$ of Definition~\ref{def:FRR}, meaning that \(\widetilde{\Sys}_1 \preceq_{\widetilde{R}}^{\FRR} \Sys_2\).\\
    $\bf{(2)\Leftrightarrow (3)}$: Directly follows from Theorem~\ref{th:FRR_concretization}.

    In addition, Theorem~\ref{th:FRR_concretization} implies
    \begin{equation}\label{eq:demo_inclusion}
        \B((\Cont_2 \circ \widetilde{R}) \times \widetilde{\Sys}_1) \subseteq \widetilde{R}^{-1}(\B(\Cont_2 \times \Sys_2)).
    \end{equation}
    Let \(\seq{x}_1 \in \B(\Cont_1 \times \Sys_1)\) be defined on \([0;N[\), where \(N \in \N \cup \{\infty\}\). Then, by Lemma~\ref{lem:equivalence_concrete_traj}, there exists a map \(\seq{z}_1\) such that \(\seq{\widetilde{x}}_1 \in \B((\Cont_2 \circ \widetilde{R}) \times \widetilde{\Sys}_1)\) with \(\widetilde{x}_1(k) = (x_1(k), z_1(k))\) for all \(k \in [0;N[\).
    Then, by~\eqref{eq:demo_inclusion}, there exists a map \(\seq{x}_2 \in \B(\Cont_2 \times \Sys_2)\) such that \(x_2(k) \in \widetilde{R}((x_1(k), z_1(k)))\) for all \(k \in [0;N[\).
    By Lemma~\ref{lem:relation_inclusion}, since $R$ satisfies~\eqref{eq:Rtilde_to_R}, we can conclude that \(x_2(k) \in R(x_1(k))\) for all \(k \in [0;N[\), i.e., \(\seq{x}_2 \in R(\seq{x}_1)\). Thus \(\B(\Cont_1 \times \Sys_1) \subseteq R^{-1}(\B(\Cont_2 \times \Sys_2))\). \hfill\hfill\qed
\end{pf}

\begin{figure*}
    \centering
    \scalebox{0.96}{     
     \begin{minipage}{\textwidth}
        \centering
         \begin{subfigure}{0.33\textwidth}
            \centering 
                 \includegraphics[width=1.0\textwidth, trim=1.6cm 0 1.6cm 0, clip]{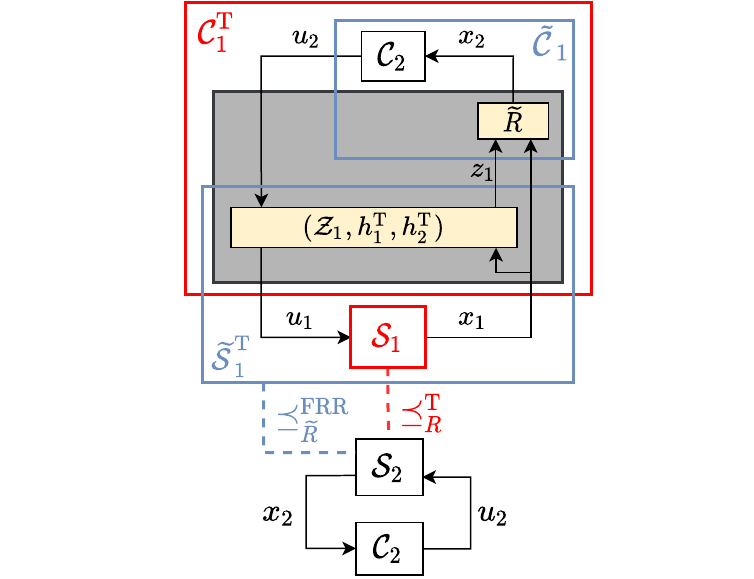}
            \caption{Concretization procedure.}
            \label{fig:augmented_system_bis}
        \end{subfigure}
        \begin{subfigure}{0.33\textwidth}
            \centering
            \includegraphics[width=1.0\textwidth, height=0.3\textheight, keepaspectratio, trim=115 0 95 0, clip]{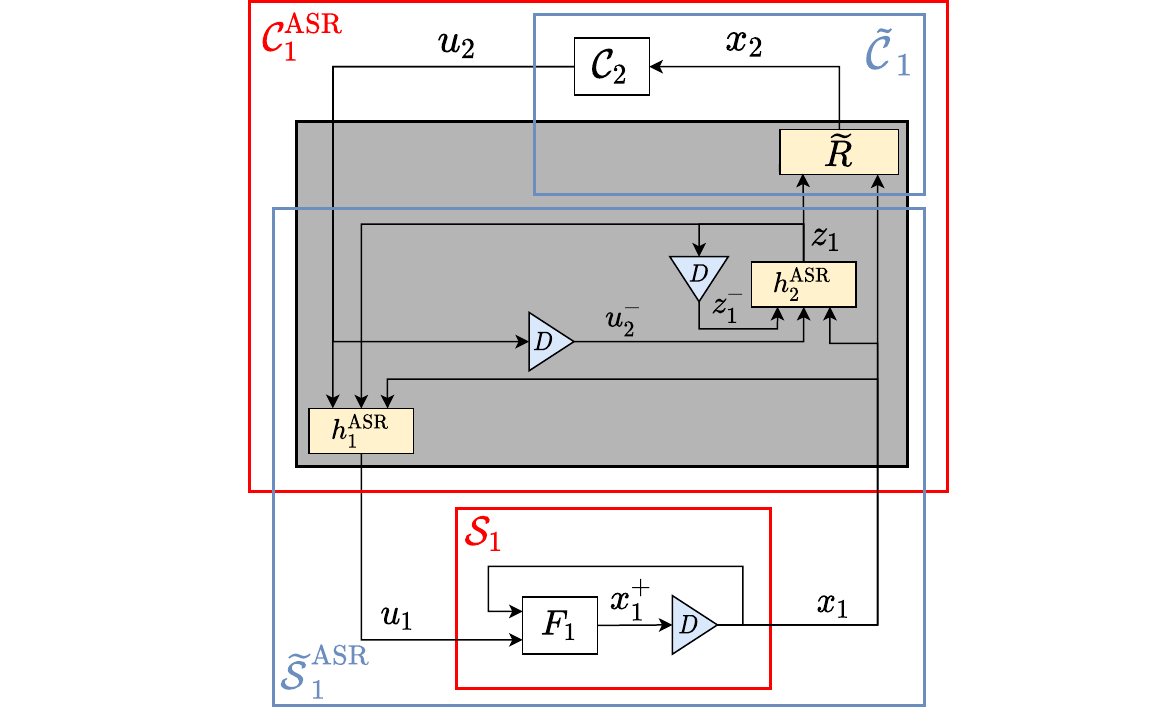}
            \caption{$\ASR$.}
            \label{fig:ASR_architecture_2}
        \end{subfigure}
        \begin{subfigure}{0.33\textwidth}
            \centering
            \includegraphics[width=1.0\textwidth, height=0.3\textheight, keepaspectratio, trim=115 0 95 0, clip]{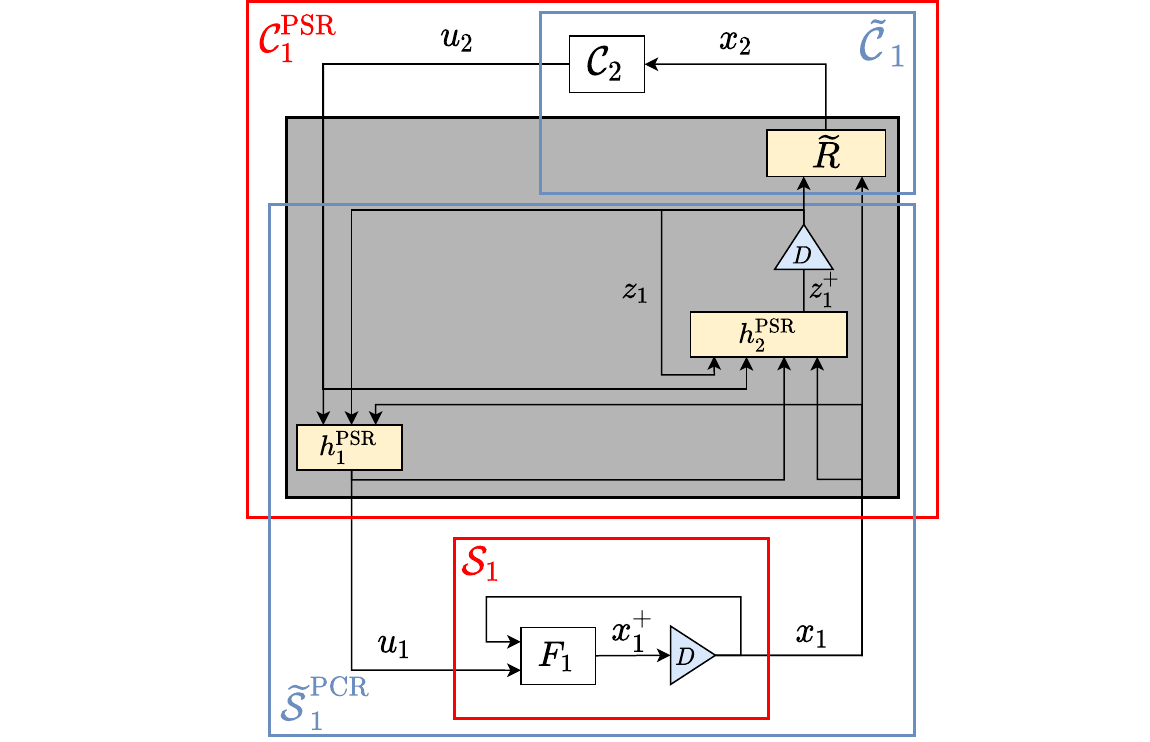}
            \caption{$\ASRB$.}
            \label{fig:ASRB_architecture_2}
        \end{subfigure}\\
        \begin{subfigure}{0.33\textwidth}
            \centering
            \includegraphics[width=1.0\textwidth, height=0.3\textheight, keepaspectratio, trim=115 0 95 0, clip]{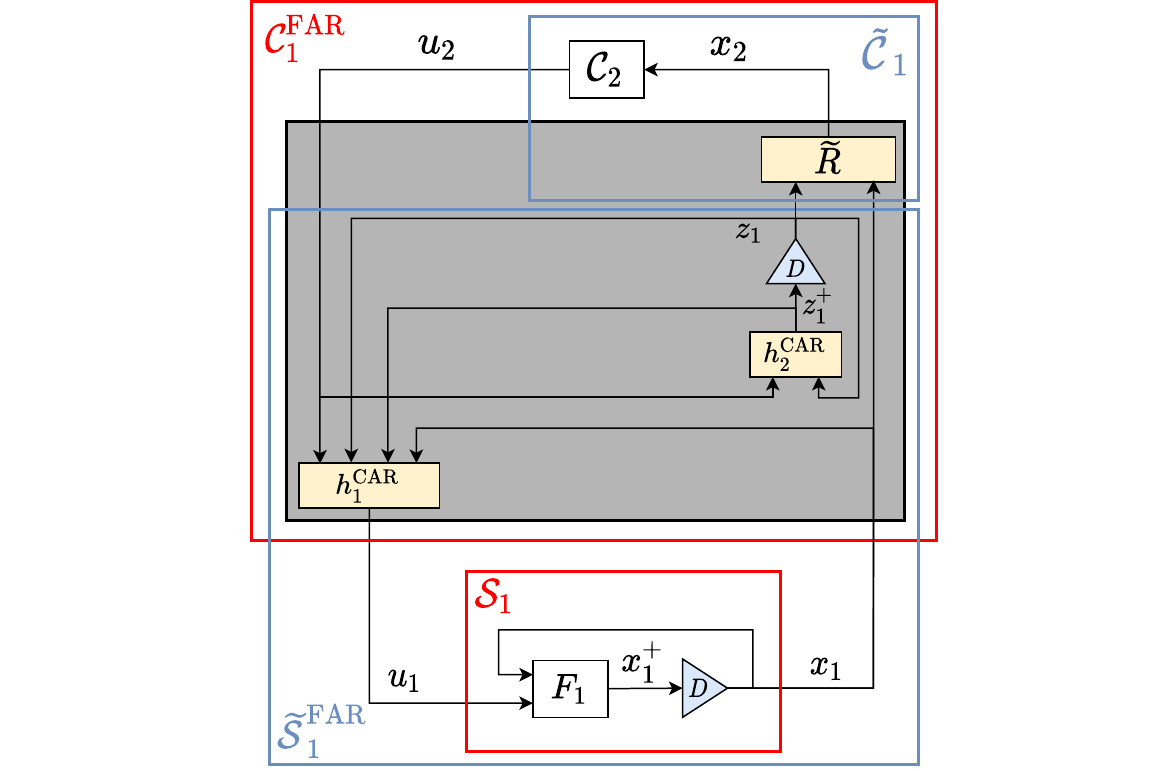}
            \caption{$\ASRBB$.}
            \label{fig:ASRBB_architecture_2}
        \end{subfigure}
        \begin{subfigure}{0.33\textwidth}
            \centering
            \includegraphics[width=1.0\textwidth, height=0.3\textheight, keepaspectratio, trim=105 0 100 0, clip]{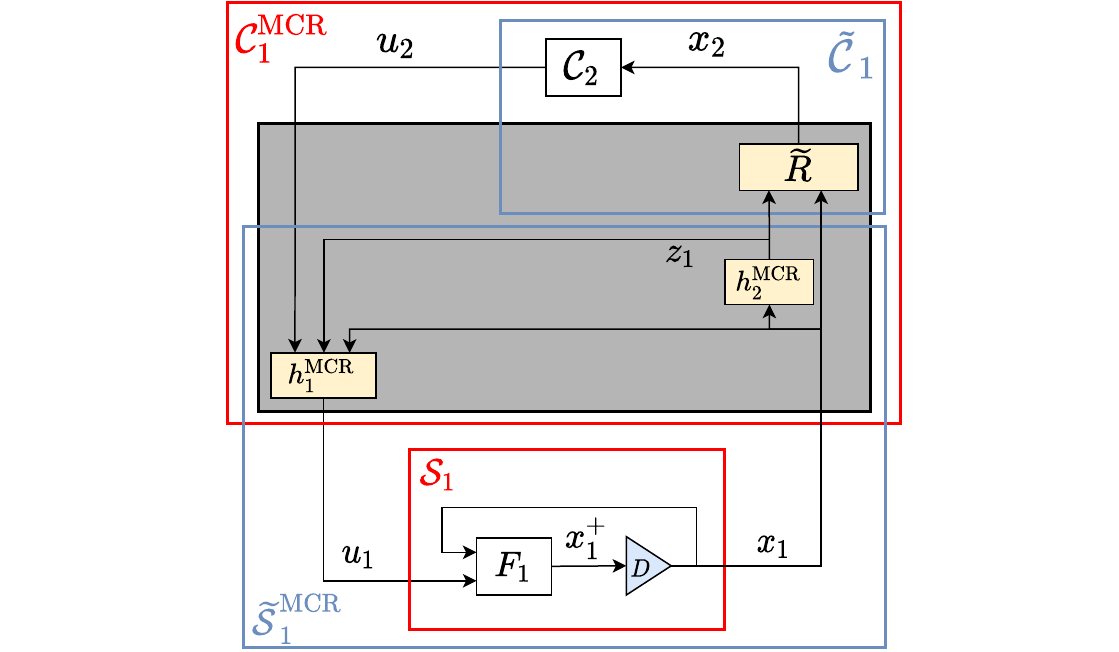}
            \caption{$\MCR$.}
            \label{fig:MCR_architecture_2}
        \end{subfigure}
        \begin{subfigure}{0.33\textwidth}
            \centering
            \includegraphics[width=1.0\textwidth, height=0.3\textheight, keepaspectratio, trim=120 0 100 0, clip]{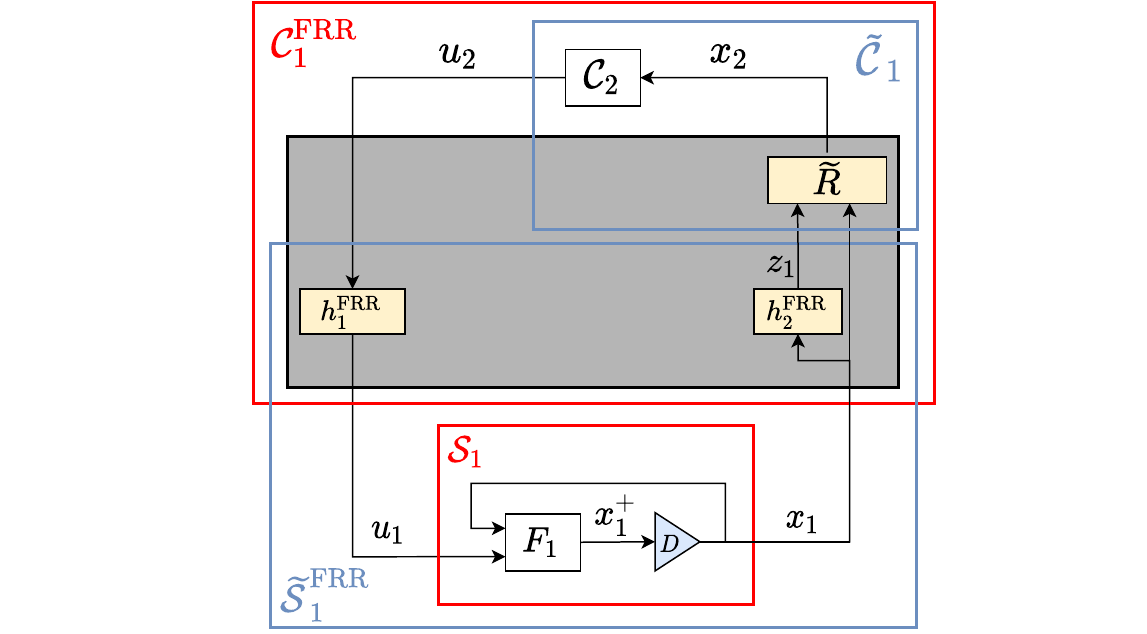}
            \caption{$\FRR$.}
            \label{fig:FRR_architecture_2}
        \end{subfigure}
        \end{minipage}
    } 
    \caption{Concretization of an abstract controller \(\Cont_2\) into a concrete controller \(\Cont_1^{\T}\) using the interface \((\set{Z}_1, h_1^{\T}, h_2^{\T}, \widetilde{R})\) for the simulation relation $\T$. (a) Depicts the interfaced system \(\widetilde{\Sys}_1^{\T}\), where \(\widetilde{\Sys}_1 \preceq_{\widetilde{R}}^{\FRR} \Sys_2\), with controller \(\widetilde{\Cont}_1 = \Cont_2 \circ \widetilde{R}\)~\eqref{eq:concrete_controller_FRR_def}. Dotted lines show the simulation relations between systems. The remaining subfigures illustrate implementations (Definition~\ref{def:interface_relations_specific_augmented}) for \(\T \in \{\ASR, \ASRB, \ASRBB, \MCR, \FRR\}\).}
    \label{fig:architectures_2}
\end{figure*}

In the next definition, we construct for each simulation relation \(\T\), the corresponding interface structure, solely determined by the variables \(\nu_1\) and \(\nu_2\) that define the domains of the functions \(h_1^{\T}\) and \(h_2^{\T}\), as illustrated in Figure~\ref{fig:architectures_2}.
\begin{definition}\label{def:interface_relations_specific_augmented}
    For each \(\T \in \{\ASR,\allowbreak\ \ASRB,\allowbreak\ \ASRBB,\allowbreak\ \MCR,\allowbreak\ \FRR\}\), we associate a tuple \((\set{Z}_1, h_1^{\T}, h_2^{\T}, \widetilde{R})\), where the functions \(h_1^{\T}\) and \(h_2^{\T}\) take as arguments the variables \(\nu_1\) and \(\nu_2\), respectively, defined as follows
    \begin{align*}
        \ASR:\ & \nu_1 = (z_1, u_2, x_1), \ 
        \nu_2 =(z_1, u_2, x_1^+); \\ 
        \ASRB:\ & \nu_1 = (z_1, u_2, x_1), \ 
        \nu_2 = (z_1, u_2, x_1, u_1); \\
        \ASRBB:\ & \nu_1 = (z_1, u_2, x_1, z_1^+), \ 
        \nu_2 = (z_1, u_2); \\
        \MCR:\ & \nu_1 = (z_1, u_2, x_1), \ 
        \nu_2 = (x_1^+); \\
        \FRR:\ & \nu_1 = (u_2), \ 
        \nu_2 = (x_1^+). 
    \end{align*}
\end{definition}
The arguments $\nu_1$ and $\nu_2$ are directly inferred from the logical structure of the formula defining the relation \(\T\). More precisely, \(\nu_1\) comprises the variables (with \(x_2\) and \(x_2^+\) replaced by \(z_1\) and \(z_1^+\)) that appear before the existential quantifier for \(u_1\), while \(\nu_2\) includes those appearing before the quantifier for \(x_2^+\). Intuitively, this reflects the role of \(h_1^{\T}\) in selecting the concrete input \(u_1\), and \(h_2^{\T}\) in determining a consistent successor \(z_1^+\), to satisfy the formula defining the relation.
For example, in the case of \(\ASR\), the quantified variables in~\eqref{eq:ASR_def} are \(x_2\), \(u_2\), \(x_1\), \(u_1\), \(x_1^+\), and \(x_2^+\). Thus, \(u_1\) depends on \(\nu_1 = (z_1, u_2, x_1)\), and \(x_2^+\) depends on \(\nu_2 = (z_1, u_2, x_1^+)\), which defines the interface of \(\ASR\), as illustrated in Figure~\ref{fig:ASR_architecture_2}.

\begin{figure}
    \centering
    \includegraphics[width=0.35\textwidth, trim=90 0 0 0, clip]{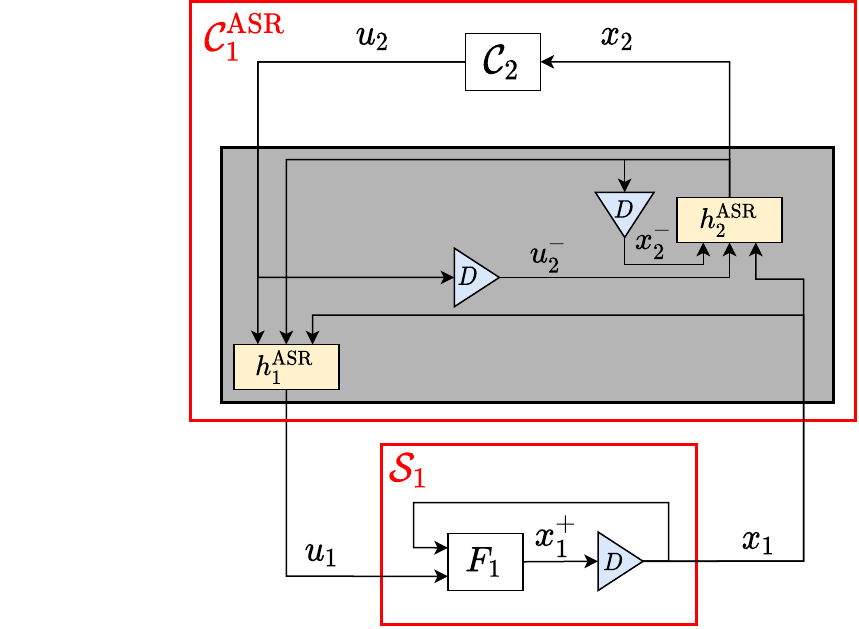}
    \caption{Simplified control architecture of Figure~\ref{fig:ASR_architecture_2} for \(\ASR\), based on the specific interface of Definition~\ref{def:interface_relations_specific}.
    }
    \label{fig:ASR_architecture_1}%
\end{figure}

Having identified in Definition~\ref{def:interface_relations_specific_augmented} the \emph{structure} of the interface for each simulation relation \(\T\), we now provide a specific \emph{implementation} of \(h_1^{\T}\) and \(h_2^{\T}\) for the specific case where \(\set{Z}_1 = \set{X}_2\), i.e., \(z_1 = x_2\) and \(z_1^+ = x_2^+\).
\begin{definition}\label{def:interface_relations_specific}
    For each \(\T \in \{\ASR,\allowbreak\ \ASRB,\allowbreak\ \ASRBB,\allowbreak\ \MCR,\allowbreak\ \FRR\}\), given two simple systems \(\Sys_1\) and \(\Sys_2\) as in~\eqref{eq:simple_sys} and a relation \(R \subseteq \set{X}_1 \times \set{X}_2\), we define the tuple \((\set{Z}_1, h_1^{\T}, h_2^{\T}, \widetilde{R})\) as follows.
    We set \(\set{Z}_1 = \set{X}_2\), \(h_1^{\T} = I_R^{\T}\) with $I_R^{\T}$ defined in~\eqref{eq:interface_abstract_concrete_input}, and
    \begin{align*}
        h_2^{\ASR}(x_2,u_2,x_1^+)=& F_2(x_2,u_2)\cap R(x_1^+),\\ 
        h_2^{\ASRB}(x_2,u_2,x_1,u_1)=& F_2(x_2,u_2)\cap \{x_2^+\in \set{X}_2:\\
        &F_1(x_1, u_1)\subseteq R^{-1}(x_2^+)\},\\
        h_2^{\ASRBB}(x_2,u_2) =& F_2(x_2,u_2)\cap \{x_2^+\in \set{X}_2:\\
        \forall x_1\in R^{-1}(x_2) \exists u_1\in& \set{U}_{\Sys_1}(x_1): 
        F_1(x_1, u_1)\subseteq R^{-1}(x_2^+)\},\\
        h_2^{\FRR}(x_1^+) =& h_2^{\MCR}(x_1^+)= R(x_1^+),
    \end{align*}
and define the lifted relation as
\begin{equation}\label{eq:R_to_Rtilde}
    \widetilde{R} = \{((x_1, x_2), x_2) \mid (x_1, x_2) \in R\}.
\end{equation}
\hfill $\triangle$
\end{definition}
In the setting of Definition~\ref{def:interface_relations_specific}, the function \(h_1^{\T}\) selects a concrete input \(u_1\) such that the successor \(x_1^+ \in F_1(x_1, u_1)\) remains related, i.e., $(x_1^+,x_2^+)\in R$, to an abstract successor \(x_2^+ \in F_2(x_2, u_2)\), identified by \(h_2^{\T}\). 
The resulting control architecture for \(\ASR\) is illustrated in Figure~\ref{fig:ASR_architecture_1}, which simplifies the general architecture for $\ASR$ of Definition~\ref{def:interface_relations_specific_augmented} shown in Figure~\ref{fig:ASR_architecture_2}.

\begin{definition}\label{as:ASRB}
    A relation $\widetilde{R}\subseteq (\set{X}_1\times \set{Z}_1)\times \set{X}_2$ satisfies the \emph{common quantization assumption} if
    $\forall x_1, x_1'\in \set{X}_1,\ z_1\in\set{Z}_1$ such that $\widetilde{R}((x_1, z_1)) \ne \myemptyset$ and $\widetilde{R}((x_1', z_1)) \ne \myemptyset$ implies $\widetilde{R}(x_1, z_1)\cap \widetilde{R}((x_1', z_1))\ne \myemptyset.$
    \hfill $\triangle$
\end{definition}

The following theorem establishes that the existence of a simulation relation between a system and its abstraction is equivalent to the existence of an interface, i.e., of a feedback refinement relation between the corresponding interfaced system and the abstraction.
\begin{theorem}\label{th:augmented_system}
    Given \(\T \in \{\ASR, \ASRB, \ASRBB, \MCR, \FRR\}\) and the simple systems \(\Sys_1\) and \(\Sys_2\) in~\eqref{eq:simple_sys}, the following propositions hold.
    \begin{enumerate}
        \item[(1)] If there exists \(R \subseteq \set{X}_1 \times \set{X}_2\) such that \(\Sys_1 \preceq_R^{\T} \Sys_2\), then \((\set{Z}_1, h_1^{\T}, h_2^{\T}, \widetilde{R})\) given in Definition~\ref{def:interface_relations_specific} is an interface from $\Sys_2$ to $\Sys_1$, with $\widetilde{R}$ satisfying Definition~\ref{as:ASRB}.
        \item[(2)] Conversely, if there exists an interface \((\set{Z}_1, h_1^{\T}, h_2^{\T}, \widetilde{R})\) 
        as given in Definition~\ref{def:interface_relations_specific_augmented}, where \(\widetilde{R}\) satisfies Definition~\ref{as:ASRB}, then \(\Sys_1 \preceq_{R}^{\T} \Sys_2\) with $R$ in~\eqref{eq:Rtilde_to_R}.
    \end{enumerate}
\end{theorem}
\begin{pf}
    See Appendix~\ref{pf:augmented_system}. \hfill\hfill\qed
\end{pf}
The following corollary of Theorem~\ref{th:augmented_system} shows that the existence of a simulation relation between the original system and its abstraction is not only sufficient to ensure the simple closed-loop structure in Figure~\ref{fig:concretization_augmented} (right), but also necessary. It also confirms that the specific interface implementation given in Definition~\ref{def:interface_relations_specific} is valid.
\begin{corollary}\label{cor:concrete_controller_augmented}
    Let \(\T \in \{\ASR,\ASRB,\ASRBB,\MCR,\FRR\}\) and two simple systems \(\Sys_1\) and \(\Sys_2\) as in~\eqref{eq:simple_sys}.
    The following propositions are equivalent.
    \begin{enumerate}
        \item[(1)] There exists $R\subseteq \set{X}_1\times \set{X}_2$ such that $\Sys_1\preceq_R^{\T} \Sys_2$;
        \item[(2)] There exists an interface $(\set{Z}_1,h_1^{\T},h_2^{\T},\widetilde{R})$ from $\Sys_2$ to $\Sys_1$ in Definition \ref{def:interface_relations_specific_augmented} with $\widetilde{R}$ satisfying Definition~\ref{as:ASRB};
    \end{enumerate}
    with $R$ and $\widetilde{R}$ satisfying~\eqref{eq:Rtilde_to_R}.
    Additionally, for any controllers $\Cont_2$ f.c. with $\Sys_2$, the concretized controller $\Cont_1^{\T}$ (Definition\ref{def:concretized_controller}) associated with the interface in Definition~\ref{def:interface_relations_specific} satisfies
    $\B(\Cont_1^{\T} \times \Sys_1) \subseteq R^{-1}(\B(\Cont_2 \times \Sys_2)).$
\end{corollary}
\begin{pf}
    It directly follows from Theorem~\ref{th:concrete_controller_augmented} and Theorem~\ref{th:augmented_system}.
\end{pf}

\begin{rem}
    The assumption that \(\widetilde{R}\) satisfies Definition~\ref{as:ASRB} in Theorem~\ref{th:augmented_system} and Corollary~\ref{cor:concrete_controller_augmented} is only required for \(\T \in \{\ASRB, \ASRBB\}\), and is automatically satisfied by the relation \(\widetilde{R}\) in~\eqref{eq:R_to_Rtilde}, as stated in~Theorem~\ref{th:augmented_system}.
    \hfill $\triangle$
\end{rem}
Finally, we define in our control formalism (Section~\ref{sec:control_framework}) the concretized controllers $\Cont_1^{\T}$ obtained by connecting \(\Cont_2\) to the interfaces \((\set{Z}_1, h_1^{\T}, h_2^{\T}, \widetilde{R})\) of Definition~\ref{def:interface_relations_specific_augmented}, as illustrated in Figure~\ref{fig:architectures_2}.
\begin{proposition}\label{def:concretized_controller_formalism}
Given \(\Sys_1\) and \(\Sys_2\) as in~\eqref{eq:simple_sys}, the interfaces from Definition~\ref{def:interface_relations_specific_augmented}, and a controller \(\Cont_2\) as in~\eqref{eq:cont}, let \(\widetilde{\Cont}_1 \defEqual \Cont_2 \circ \widetilde{R}\). Then, the controllers \(\Cont_1^{\T} = (\set{X}_{\Cont_1^{\T}}, \set{X}_1, \set{V}_{\Cont_1^{\T}}, \set{U}_1, F_{\Cont_1^{\T}}, H_{\Cont_1^{\T}})\) satisfy Definition~\ref{def:concretized_controller}.
    {\small
    \begin{align*}
        \bullet\ &\set{X}_{\Cont_1^{\ASR}} = \set{X}_{\widetilde{\Cont}_1} \times \set{Z}_1 \times \set{U}_2,\ \set{V}_{\Cont_1^{\ASR}} = \set{V}_{\widetilde{\Cont}_1} \times \set{Z}_1 \times \set{U}_2,\\
        &F_{\Cont_1^{\ASR}}((x_{\widetilde{\Cont}_1},z_1^-,u_2^-),(v_{\widetilde{\Cont}_1},z_1,u_2)) =  
        \{(x_{\widetilde{\Cont}_1}^+,z_1,u_2)\mid\\ &\hspace{5.08cm}x_{\widetilde{\Cont}_1}^+ \in F_{\widetilde{\Cont}_1}(x_{\widetilde{\Cont}_1}, v_{\widetilde{\Cont}_1})\},\\
        &H_{\Cont_1^{\ASR}}((x_{\widetilde{\Cont}_1},z_1^-,u_2^-),x_1) = 
        \{(u_1,(v_{\widetilde{\Cont}_1},z_1,u_2))\mid\\ 
        &\hspace{0.68cm}z_1 \in h_2^{\ASR}(z_1^-,u_2^-,x_1),
        (u_2,v_{\widetilde{\Cont}_1}) \in H_{\widetilde{\Cont}_1}(x_{\widetilde{\Cont}_1}, (x_1,z_1)),\\
        &\hspace{4.85cm}u_1 \in h_1^{\ASR}(z_1,u_2,x_1)\}.\\
        \bullet\ &\set{X}_{\Cont_1^{\ASRB}} = \set{X}_{\widetilde{\Cont}_1} \times \set{Z}_1, \ \set{V}_{\Cont_1^{\ASRB}} =\set{V}_{\widetilde{\Cont}_1} \times \set{Z}_1,\\
        &F_{\Cont_1^{\ASRB}}((x_{\widetilde{\Cont}_1},z_1),(v_{\widetilde{\Cont}_1},z_1^+)) = \{(x_{\widetilde{\Cont}_1}^+,z_1^+)
        \mid \\
        &\hspace{5.08cm}x_{\widetilde{\Cont}_1}^+\in F_{\widetilde{\Cont}_1}(x_{\widetilde{\Cont}_1}, v_{\widetilde{\Cont}_1})\},\\
        &H_{\Cont_1^{\ASRB}}((x_{\widetilde{\Cont}_1},z_1),x_1) = \ 
        \{(u_1,(v_{\widetilde{\Cont}_1},z_1^+))\mid\\ 
        &\hspace{0.85cm}(u_2,v_{\widetilde{\Cont}_1})\in H_{\widetilde{\Cont}_1}(x_{\widetilde{\Cont}_1}, (x_1,z_1)), u_1\in h_1^{\ASRB}(z_1,u_2,x_1),\\
        &\hspace{4.37cm}z_1^+\in h_2^{\ASRB}(z_1,u_2,x_1,u_1)\}.\\
        \bullet\ & \set{X}_{\Cont_1^{\ASRBB}} = \set{X}_{\widetilde{\Cont}_1} \times \set{Z}_1, \ \set{V}_{\Cont_1^{\ASRBB}}, =\set{V}_{\widetilde{\Cont}_1} \times \set{Z}_1,\\
        &F_{\Cont_1^{\ASRBB}}((x_{\widetilde{\Cont}_1},z_1),(v_{\widetilde{\Cont}_1},z_1^+)) = \{(x_{\widetilde{\Cont}_1}^+,z_1^+)\mid \\
        &\hspace{5.08cm}x_{\widetilde{\Cont}_1}^+\in F_{\widetilde{\Cont}_1}(x_{\widetilde{\Cont}_1}, v_{\widetilde{\Cont}_1})\},\\
        &H_{\Cont_1^{\ASRBB}}((x_{\widetilde{\Cont}_1},z_1),x_1) = \ 
        \{(u_1,(v_{\widetilde{\Cont}_1},z_1^+))\mid \\
        &\hspace{1.24cm}(u_2,v_{\widetilde{\Cont}_1})\in H_{\widetilde{\Cont}_1}(x_{\widetilde{\Cont}_1}, (x_1,z_1)),
        z_1^+\in h_2^{\ASRBB}(z_1,u_2), \\
        &\hspace{4.135cm}\hspace{0.2cm} 
        u_1\in h_1^{\ASRBB}(z_1,u_2,x_1,z_1^+)\}.\\
        \bullet\ & \set{X}_{\Cont_1^{\MCR}} = \set{X}_{\widetilde{\Cont}_1},\ \set{V}_{\Cont_1^{\MCR}} =\set{V}_{\widetilde{\Cont}_1},\\
        &F_{\Cont_1^{\MCR}}(x_{\widetilde{\Cont}_1}, v_{\widetilde{\Cont}_1}) = F_{\widetilde{\Cont}_1}(x_{\widetilde{\Cont}_1}, v_{\widetilde{\Cont}_1}),\\
        &H_{\Cont_1^{\MCR}}(x_{\widetilde{\Cont}_1},x_1) = 
        \{(u_1, v_{\widetilde{\Cont}_1}) \mid 
        z_1\in h_2^{\MCR}(x_1),\\
        &\hspace{0.60cm} (u_2,v_{\widetilde{\Cont}_1})\in H_{\widetilde{\Cont}_1}(x_{\widetilde{\Cont}_1}, (x_1,z_1)), u_1\in h_1^{\MCR}(z_1,u_2,x_1)\}.\\
        \bullet\ & \set{X}_{\Cont_1^{\FRR}} = \set{X}_{\widetilde{\Cont}_1}, \ \set{V}_{\Cont_1^{\FRR}} = \set{V}_{\widetilde{\Cont}_1},\\
        &F_{\Cont_1^{\FRR}}(x_{\widetilde{\Cont}_1}, v_{\widetilde{\Cont}_1}) = F_{\widetilde{\Cont}_1}(x_{\widetilde{\Cont}_1}, v_{\widetilde{\Cont}_1}),\\
        &H_{\Cont_1^{\FRR}}(x_{\widetilde{\Cont}_1},x_1) = \{(u_1, v_{\widetilde{\Cont}_1}) \mid z_1 \in h_2^{\FRR}(x_1), \\
        &\hspace{1.54cm} (u_2, v_{\widetilde{\Cont}_1}) \in H_{\widetilde{\Cont}_1}(x_{\widetilde{\Cont}_1}, (x_1,z_1)), u_1 \in h_1^{\FRR}(u_2)\}.
    \end{align*}
    }
\end{proposition}
\begin{pf}
    This follows from Theorem~\ref{th:augmented_system} and Definition~\ref{def:feedback_composition}.
    \hfill\qed
\end{pf}
Proposition~\ref{def:concretized_controller_formalism} provides an implementation of the controller \(\Cont_1^{\ASR}\) of Theorem~\ref{th:ASR_concretization_property} in our control framework. 
In addition, we recover the controller $\Cont_1^{\FRR}$ of Theorem~\ref{th:FRR_concretization} using our framework (Proposition~\ref{def:concretized_controller_formalism}) in the setting of Definition~\ref{def:interface_relations_specific}. Indeed, in that case
$H_{\Cont_1^{\FRR}}(x_{\Cont_2}, x_1) = \{(u_2, v_{\Cont_2}) \mid x_2 \in R(x_1), (u_2, v_{\Cont_2}) \in H_{\Cont_2}(x_{\Cont_2}, x_2)\} = H_{\Cont_2}(x_{\Cont_2}, R(x_1)),$ establishing that \(\Cont_1^{\FRR} = \Cont_2 \circ R\) as in~\eqref{eq:concrete_controller_FRR_def}. Consequently, for \(\T = \FRR\), the architecture in Figure~\ref{fig:FRR_architecture_2} reduces to that in Figure~\ref{fig:concretization_augmented} (left), and Corollary~\ref{cor:concrete_controller_augmented} recovers Theorem~\ref{th:FRR_concretization}.
Therefore, Corollary~\ref{cor:concrete_controller_augmented} generalizes the concretization result of Theorem~\ref{th:FRR_concretization} to simulation relations beyond the feedback refinement relation.

To summarize, we introduced a general framework for characterizing simulation relations through their associated control architectures by defining a generic interface structure based on two set-valued functions, \(h_1^{\T}\) and \(h_2^{\T}\), which systematically capture the concretization mechanism for any relation refining the \(\ASR\).
Importantly, the \emph{arguments} of the functions \(h_1^{\T}\) and \(h_2^{\T}\)—which define the block representation of the interface (see Figure~\ref{fig:architectures_2})—depend only on the type of relation \(\T\), and not on the specific systems \(\Sys_1\) and \(\Sys_2\) (see Definition~\ref{def:interface_relations_specific_augmented}). While the \emph{implementation} of these blocks does depend on the systems involved (see Definition~\ref{def:interface_relations_specific}), it remains independent of the abstract controller \(\Cont_2\). This makes the interface specification-agnostic, enabling a plug-and-play concretization procedure as outlined in the introduction.

\section{Properties}\label{sec:discussion}

This section analyzes how each previously introduced simulation relation affects the abstraction structure, controller complexity, and achievable performance.  
These aspects are illustrated throughout using the simple one-dimensional running example shown in Figure~\ref{fig:example_properties}.


\subsection{The concretization complexity issue}\label{sec:discussion_concretization_complexity_issue}

The implementations of $h_2^{\ASR}$, $h_2^{\ASRB}$, and $h_2^{\ASRBB}$ in Definition~\ref{def:interface_relations_specific} require online evaluation of $F_2$ of $\Sys_2$.  
Given the typically large state space of $\Sys_2$, this becomes computationally demanding, as the concretized controllers $\Cont_1^{\ASR}$, $\Cont_1^{\ASRB}$, and $\Cont_1^{\ASRBB}$ must store abstract variables and query the abstraction at runtime.  
In contrast, $h_2^{\MCR}$ depends solely on the quantizer $R$, not on $F_2$, making the $\MCR$ concretization considerably simpler.

In the system described in Figure~\ref{fig:example_properties}, the optimal abstract policy for $\ASR$, achieves a worst-case cost of~$3$ transitions ($q_0\!\to\!q_2\!\to\!q_3\!\to\!q_4$).  
However, to ensure that a concrete successor $x_1^+$ is quantized into $q_2$—and not mistakenly into $q_1$—the concretized controller must compute
$x_2^+ \in h_2^{\ASR}(x_2,u_2,x_1^+) = F_2(x_2,u_2)\cap R(x_1^+),$
thus storing $(x_2,u_2)$ and accessing $F_2$ online.  
This illustrates the \emph{concretization complexity} discussed above.

Under $\MCR$, the abstract transition is conservatively enlarged to include both $q_1$ and $q_2$ in $F_2(q_0,\cdot)$, which allows the concretized controller to compute $h_2^{\MCR}(x_1^+)=R(x_1^+)$ directly. This removes the need for storing abstract variables $x_2$ and $u_2$, and for evaluating the abstract transition map $F_2$ during online execution.  
The concretized controller $\Cont_1^{\MCR}$ is therefore static, as formalized in the next proposition; however, the added non-determinism increases the worst-case cost from~$3$ to~$4$ ($q_0\!\to\!q_1\!\to\!q_2\!\to\!q_3\!\to\!q_4$).




Even when the abstract controller $\Cont_2$ is static (i.e., $\set{X}_{\Cont_2}=\{0\}$), the concretized controllers $\Cont_1^{\ASR}$, $\Cont_1^{\ASRB}$, and $\Cont_1^{\ASRBB}$ remain dynamic, as they must store abstract variables—represented by the delay blocks in Figure~\ref{fig:architectures_2}.  
In contrast, the concretized controller $\Cont_1^{\MCR}$ introduces no additional state (no delay block in Figure~\ref{fig:MCR_architecture_2}).  
Hence, if $\Cont_2$ is static, $\Cont_1^{\MCR}$ is also static, as formally stated in the following proposition.




\begin{proposition}\label{prop:static}
Given two simple systems \( \Sys_1 \) and \( \Sys_2 \) as in~\eqref{eq:simple_sys}, a relation \( R \subseteq \set{X}_1 \times \set{X}_2 \), and a controller \( \Cont_2 \) f.c. with \( \Sys_2 \), if \( \Cont_2 \) is static, then the corresponding concretized controller \( \Cont_1^{\MCR} \) (as defined in Proposition~\ref{def:concretized_controller_formalism}) is also static.
\end{proposition}
\begin{pf}
This follows directly from Proposition~\ref{def:concretized_controller_formalism}, where \( \set{X}_{\Cont_1^{\MCR}} = \set{X}_{\Cont_2} \), $\set{V}_{\Cont_1^{\MCR}}$ $= \set{V}_{\Cont_2}$, and \( F_{\Cont_1^{\MCR}} = F_{\Cont_2} \). Therefore, if \( \Cont_2 \) is static (i.e., \( \set{X}_{\Cont_2} = \{0\} \)), then \( \set{X}_{\Cont_1^{\MCR}} = \{0\} \), confirming that \( \Cont_1^{\MCR} \) is static. \hfill\qed
\end{pf}

\begin{figure}
    \centering
    \includegraphics[trim=1.78cm 0 0.5cm 0, clip,width=0.48\textwidth]{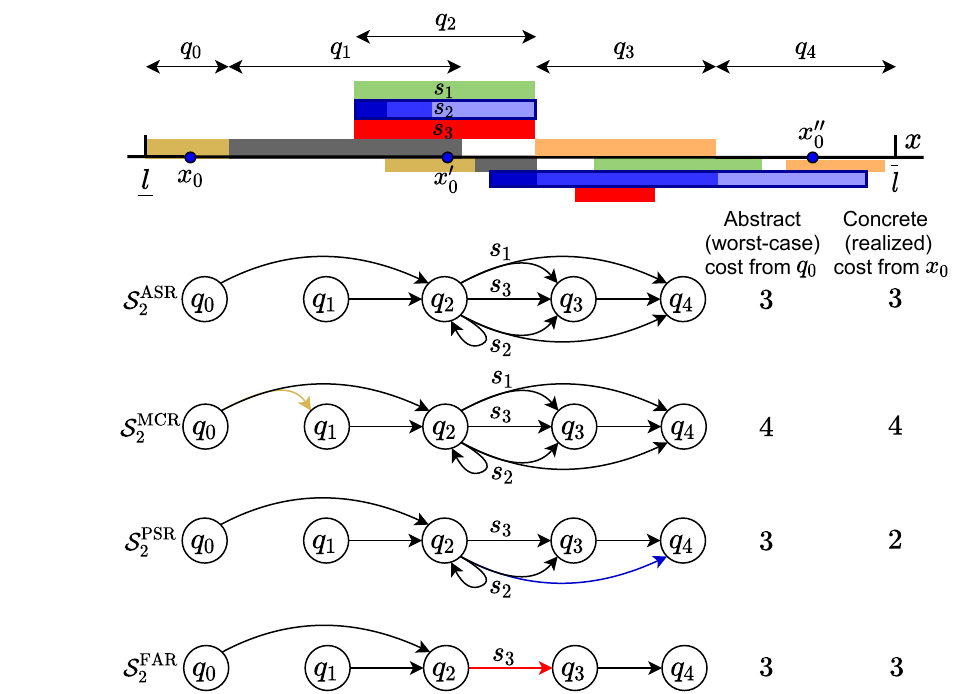}
    \caption{Running example ($1$-dimensional) used to illustrate the properties of the simulation relations.  
    The deterministic system $\Sys_1=(\set{X}_1,\set{U}_1,F_1)$ with $\set{X}_1\subseteq [\underline{l},\overline{l}]\subset\R$, is discretized into five abstract cells $q_0,\dots,q_4\in\set{X}_2$. The relation is defined by $R(x)=\{\,q\in\set{X}_2 \mid x\in q\,\}$. Cells $q_1$ and $q_2$ partially overlap, which makes $R$ non-deterministic on this region. 
    Above the $x$-axis, the intervals represent subsets of $\set{X}_1$; below, in the same color, we show their closed-loop images $F_1(x,\kappa(x))$ under local feedbacks $\kappa(\cdot)$, which define the abstract inputs (only a few are labeled $s_1,s_2,s_3\in\set{U}_2$ for clarity). The objective is to drive the system from $q_0$ to $q_4$ in the minimum number of steps.  
    The figure also reports the optimal abstract and concrete cost with initial condition $x_0\in q_0$ obtained for each abstraction $\Sys_2^{\T}$, for which it can be verified that $\Sys_1 \preceq_{R}^{\T} \Sys_2^{\T}$.}
    \label{fig:example_properties}
\end{figure}

\subsection{Corrective vs predictive control architecture}\label{sec:discussion_corrective_vs_predictive}
The architecture induced by $\ASR$ is \emph{corrective}: the controller stores the current abstract state $x_2$ and input $u_2$ to adjust the next abstract state according to
$x_2^+ \in h_2^{\ASR}(x_2,u_2,x_1^+) = F_2(x_2,u_2) \cap R(x_1^+),$
at the next time step.  
In contrast, $\ASRB$ adopts a \emph{predictive} architecture: it stores the \emph{next} abstract state $x_2^+$, determined \emph{before} applying the current input $u_1$ to the concrete system.  
This difference is reflected in the delay blocks of Figure~\ref{fig:architectures_2}.

Consider the abstract input $s_1$, $s_2$, $s_3$ available at~$q_2$.  
During offline abstract synthesis under $\ASR$ (or $\ASRB$), $s_2$ is discarded because it allows a self-loop on $q_2$, leading to a worst-case abstract cost of~$3$.  
However, $\ASRB$ enables an online adaptation: the function $h_2^{\ASRB}(x_2,u_2,x_1,u_1)$ can identify the concrete subset of $q_2$ where the trajectory lies and predict the next abstract state accordingly.  
If $x_1$ belongs to the \emph{favorable} (light-blue) subset of $q_2$, the same input $s_2$ deterministically drives the system directly to $q_4$, saving one transition (see the concrete trajectory $x_0\to x_0'\to x_0''$).  
Hence, $\ASRB$ preserves the same worst-case cost ($3$) but can \emph{improve the realized cost} online through local policy updates—at the expense of higher controller complexity, since $h_2^{\ASRB}$ must store the subdivision of $q_2$ and its associated transitions.



\subsection{Feedforward abstract control}
For $\ASRBB$, the next abstract state $x_2^+$ is determined independently of the concrete variables, including the concrete state $x_1$, in contrast with the predictive architecture of $\ASRB$ (see Figure~\ref{fig:architectures_2}).  
As a result, the abstract trajectory can be fully designed offline and tracked online.  
This construction, however, typically requires local incremental stabilizability to guarantee that the image of each abstract cell lies entirely within a unique successor.

In the running example, input $s_3$ deterministically maps all states in $q_2$ into $q_3$, yielding a deterministic abstraction. The corresponding concretized controller $\Cont_1^{\ASRBB}$ simply tracks the predefined abstract sequence $q_0\!\to\!q_2\!\to\!q_3\!\to\!q_4$ using a stabilizing feedback. The cost remains exactly~$3$, as the online abstract trajectory coincides with the offline one—illustrating the purely \emph{feedforward} nature of $\ASRBB$.


Finally, most abstraction-based control frameworks in the literature rely either on a deterministic abstract system~$\Sys_2$~\citep{kloetzer2008fully,girard2009approximately,wongpiromsarn2011tulip,mouelhi2013cosyma,asselborn2013control,egidio2022state,calbert2024smart}, or on a deterministic relation~$R$~\citep{rungger2016scots,hsu2018multi,khaled2019pfaces,apaza2020symbolic,calbert2021alternating,calbert2024dionysos} corresponding to a partition of the state space without overlaps.
According to Proposition~\ref{prop:meta_relations}, these two common settings align with simulation relations $\ASRBB$ and $\MCR$, respectively.

\section{Construction}\label{sec:Constructions}

This section formalizes the construction of abstractions corresponding to each simulation relation introduced earlier.  
Building on the qualitative discussion of Section~\ref{sec:discussion}, we now provide systematic procedures for deriving, from a continuous deterministic system, abstract models that satisfy the relations $\ASR$, $\ASRB$, $\ASRBB$, and $\MCR$.  
To this end, we consider a class of deterministic systems equipped with a \emph{growth-bound function}, and use a common state-space discretization to derive the corresponding abstract transition maps.  
This framework highlights how different relations impose distinct assumptions on the original dynamics and how these assumptions shape the resulting control architectures.

\subsection{Stability preliminaries}

A continuous function $\gamma: \R^+_0\rightarrow \R^+_0$ is said to belong to class $\mathcal{K}$ if it is strictly increasing and $\gamma(0) = 0$. 
It is said to belong to class $\mathcal{K}_{\infty}$ if it is a $\mathcal{K}$ function and $\gamma(r)\rightarrow +\infty$ when $r\rightarrow+\infty$.

We introduce a relaxed version of Lyapunov function, referred to as the \emph{growth-bound function}, to bound the distance between trajectories with different initial conditions under a specific control law. This is the discrete-time analogue of~\citet[Definition 2.3]{girard2009approximately}.
\begin{definition}\label{def:Lyap}
 A smooth function $V:\R^n\times \R^n\rightarrow \R_0^+$ is a \emph{growth-bound function} under control $\kappa:\R^n\times \R^n\times \R^\nu\rightarrow \R^{\nu}$ for system $f:\R^n\times \R^{\nu}\rightarrow\R^n$ if 
 there exist $\mathcal{K}_{\infty}$ functions $\underline{\alpha},\overline{\alpha}$ and $\rho>0$ such that $\forall x,y\in\R^n$, $u\in\R^{\nu}$
 \begin{align}
    &V(f(y,\kappa(y,x,u)), f(x,u))\le \rho V(y,x) \label{eq:lyapunov}\\
    &\underline{\alpha}(\|y-x\|_2)\le V(y,x)\le \overline{\alpha}(\|y-x\|_2).
\end{align}
We define the $\ell$-sublevel set of $V(\cdot, x)$ as
\begin{equation}\label{eq:sublevel}
    \set{S}(x, \ell) = \{x'\in \R^n \mid V(x',x)\le \ell\}.
\end{equation}
\end{definition}
When $\rho<1$, the system is said to be \emph{incrementally stabilizable} meaning that, under a suitable feedback, all trajectories converge to a common reference trajectory regardless of their initial conditions.

\subsection{System definition and grid discretization}
We study a deterministic simple system $\Sys_1 = (\set{X}_1, \set{U}_1, F_1)$, where $\set{X}_1 \subseteq \R^n$ is a compact set, $\set{U}_1 \subseteq \R^{\nu}$, and the transition map $F_1(x,u) = \{f(x,u)\}$ is defined by a continuous nonlinear function $f: \set{X}_1 \times \set{U}_1 \rightarrow \set{X}_1$.
We impose stabilizability assumptions on $f$ by assuming the existence of a growth-bound function $V$ under control $\kappa$ (Definition~\ref{def:Lyap}) with a contraction factor $\rho$.
We will make the following supplementary assumption on the function $V$: there exists a $\mathcal{K}_{\infty}$ function $\gamma$ such that
\begin{equation}\label{eq:as_sup}
    \forall x,y,z\in \R^n :\ V(x, y) - V(x,z) \le \gamma(\|y-z\|_2).
\end{equation}
While this assumption may seem quite strong, it is not restrictive provided we are interested in the dynamics of the system on a compact subset of the state space $\R^n$, as explained in~\citet[Section IV.B]{girard2009approximately}.
We illustrate the constructions on the following example.
\begin{example}\label{example:2D}
We consider a two-dimensional deterministic simple system $\Sys_1=(\set{X}_1,\set{U}_1,F_1)$ (\(\set{X}_1 \subseteq \R^2\), $F_1(x_1,u_1)=\{f(x_1,u_1)\}$) with growth-bound function \( V(x,y) = \|x-y\|_2 \), where \(\underline{\alpha}(x)=\overline{\alpha}(x)=\gamma(x)=x=\underline{\alpha}^{-1}(x)=\overline{\alpha}^{-1}(x)=\gamma^{-1}(x)\), under control \(\kappa\) for $f$ and contraction factor \(\rho\). The function \( V \) satisfies~\eqref{eq:as_sup}, which reduces to the triangle inequality of the 2-norm.
\hfill $\triangle$
\end{example}

%
The abstraction constructions rely on discretizing the state space \(\set{X}_1\) into overlapping cells aligned on a grid
\begin{equation}\label{eq:grid_definition}
[\eta \Z^n] = \{c \in \R^n \mid \exists_{k \in \Z^n} \ \forall_{i \in [1;n]} \ c_i = k_i \eta\}
\end{equation}
with \emph{grid parameter} \(\eta > 0\).
The abstract states correspond to the vertices of this lattice, i.e.,
\begin{equation}\label{def:abstract_state_space}
    \set{X}_2 = [\eta \Z^n] \cap \set{X}_1.
\end{equation}
Given \(\epsilon > 0\), each abstract state \(x_2 \in \set{X}_2\) is associated with the \(\epsilon\)-sublevel set of the function \(V(\cdot, x_2)\), such that
\begin{equation}\label{eq:relation}
    (x_1, x_2) \in R \ \Leftrightarrow \ V(x_1, x_2) \le \epsilon \ \Leftrightarrow \ R^{-1}(x_2) = \set{S}(x_2, \epsilon).
\end{equation}
To ensure that the relation is strict, i.e., \(R(x_1) \neq \myemptyset\) for all \(x_1 \in \set{X}_1\), we impose the condition \(\eta \le \tfrac{2}{\sqrt{n}} \overline{\alpha}^{-1}(\epsilon)\), under which Proposition~\ref{prop:strict_relation} guarantees the existence of some \(x_2 \in \set{X}_2\) such that \(x_1 \in \set{S}(x_2, \epsilon)\).

Given $x_2\in\set{X}_2$ and $u_2\in\set{U}_2$, the growth assumption~\eqref{eq:lyapunov} allows to determine an over-approximation of the attainable set of a cell $R^{-1}(x_2)$ under the control law $g_{x_2,u_2}(x) = f(x,\kappa(x, x_2, u_2))$ (see Proposition~\ref{prop:over_approx})
\begin{equation}\label{eq:over-approx}
    g_{x_2,u_2}(R^{-1}(x_2)) \subseteq \set{S}(f(x_2,u_2),\rho \epsilon).
\end{equation}
%


\begin{figure}
    \centering
    \includegraphics[trim=2.8cm 0 0 0, clip,width=0.48\textwidth]{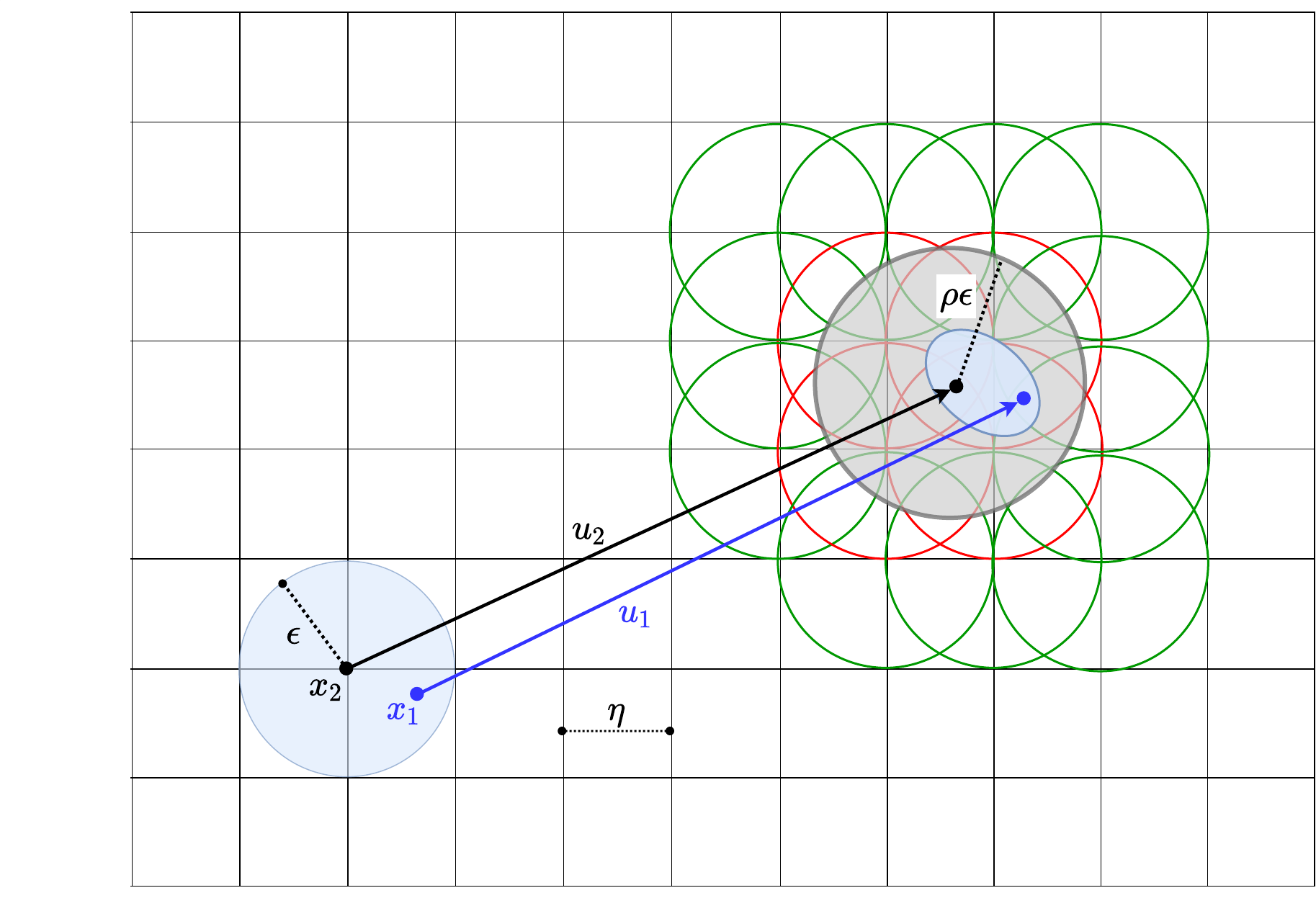}
    \caption{Illustration of $(x_2,u_2,x_1,u_1)\in R_e^{\ASR}$ and $R_e^{\MCR}$ for Example~\ref{example:2D}, for some $\rho>1$. 
    The cell $R^{-1}(x_2) = \mathcal{S}(x_2, \epsilon)$ and its image under the dynamic $F_1$ are represented in blue. The grey circle $\set{A}\defEqual \mathcal{S}(f(x_2, u_2), \rho \epsilon)$ is the over-approximation of the reachable set of $R^{-1}(x_2)$.
     The set $F_2^{\ASR}(x_2, u_2)$ just needs to contain enough cells so that their union covers the set $\set{A}$~\eqref{eq:example_ASR}, for example, only the red cells.
    While the set $F_2^{\MCR}(x_2, u_2)$ must at least contain all the cells intersecting~$\set{A}$~\eqref{eq:example_MCR}, which are shown in red and green.
    Therefore, in this example, we have $|F_2^{\ASR}(x_2,u_2)| = 4$ while $|F_2^{\MCR}(x_2,u_2)| = 15$.
    } 
    \label{fig:Example_ASR_MCR}
\end{figure}

\subsection{Abstraction construction per simulation relation}

To facilitate comparison across the different simulation relations, all abstractions are constructed over a common state space $\set{X}_2$~\eqref{def:abstract_state_space}, a common input space $\set{U}_2 \subseteq \set{U}_1$, and the same relation $R \subseteq \set{X}_1 \times \set{X}_2$~\eqref{eq:relation}.  
Hence, their differences arise solely from the definition of the transition map $F_2^{\top}$ deigned so that the abstract system $\Sys_2^{\T}=(\set{X}_2,\set{U}_2,F_2^{\T})$ satisfies $\Sys_1 \preceq_R^{\T} \Sys_2^{\T}$.

\begin{proposition}\label{prop:example_ASR}
The system $\Sys_2^{\ASR} = (\set{X}_2, \set{U}_2, F_2^{\ASR})$ with $F_2^{\ASR}$ satisfying $\forall x_2\in\set{X}_2,\ u_2\in\set{U}_{\Sys_2}(x_2)$:
\begin{equation}\label{eq:example_ASR}
     R^{-1}(F_2^{\ASR}(x_2, u_2)) \supseteq  \set{S}(f(x_2,u_2), \rho \epsilon),
\end{equation}
ensures $\Sys_1\preceq_R^{\ASR}\Sys_2^{\ASR}$. 
Similarly, the system $\Sys_2^{\MCR} = (\set{X}_2, \set{U}_2, F_2^{\MCR})$ with $F_2^{\MCR}$ satisfying $\forall x_2\in\set{X}_2,\ u_2\in\set{U}_{\Sys_2}(x_2)$:
\begin{equation}\label{eq:example_MCR}
    F_2^{\MCR}(x_2, u_2) \supseteq R(\set{S}(f(x_2,u_2), \rho \epsilon)),
\end{equation}
ensures $\Sys_1\preceq_R^{\MCR}\Sys_2^{\MCR}$. 
Additionally, one can use the specific implementations $h_1^{\ASR}(x_2,u_2,x_1) = h_1^{\MCR}(x_2,u_2,x_1) = \{\kappa(x_1,x_2,u_2)\}$. 
\end{proposition}
\begin{pf}
Let $(x_1, x_2) \in R$ and $u_2 \in \set{U}_{\Sys_2}(x_2)$, then for all $x_1^+ \in F_1(x_1, u_1)$ with $u_1 = \kappa(x_1, x_2, u_2)$, we have $x_1^+ \in \set{S}(f(x_2, u_2), \rho \epsilon)$ by~\eqref{eq:over-approx}.
\begin{itemize}
    \item Condition~\eqref{eq:example_ASR} involves $x_1^+\in \set{S}(f(x_2, u_2), \rho \epsilon) \subseteq R^{-1}(F_2^{\ASR}(x_2, u_2))$, which implies that $R(x_1^+) \cap F_2^{\ASR}(x_2, u_2) \ne \myemptyset$. This establishes~\eqref{eq:ASR_def} and that $\kappa(x_1, x_2, u_2) \in I_R^{\ASR}(x_2, u_2,x_1)$.
    \item Condition~\eqref{eq:example_MCR} involves $R(x_1^+) \subseteq F_2(x_2, u_2)$, which establishes~\eqref{eq:MCR_def} and that $\kappa(x_1, x_2, u_2) \in I_R^{\MCR}(x_2, u_2,x_1)$. \hfill\hfill\qed
\end{itemize}
\end{pf}
While $\MCR$ yields a simpler concretization than $\ASR$ (Section~\ref{sec:discussion_concretization_complexity_issue}), it increases abstraction non-determinism: 
$F_2^{\ASR}(x_2,u_2)\subseteq F_2^{\MCR}(x_2,u_2)$ (Figure~\ref{fig:Example_ASR_MCR}).  
The larger branching factor can make the abstract control problem infeasible for $\Sys_2^{\MCR}$ even when feasible for~$\Sys_2^{\ASR}$.


\begin{figure}
    \centering
    \includegraphics[width=0.48\textwidth, keepaspectratio, trim=80 0 80 0, clip]{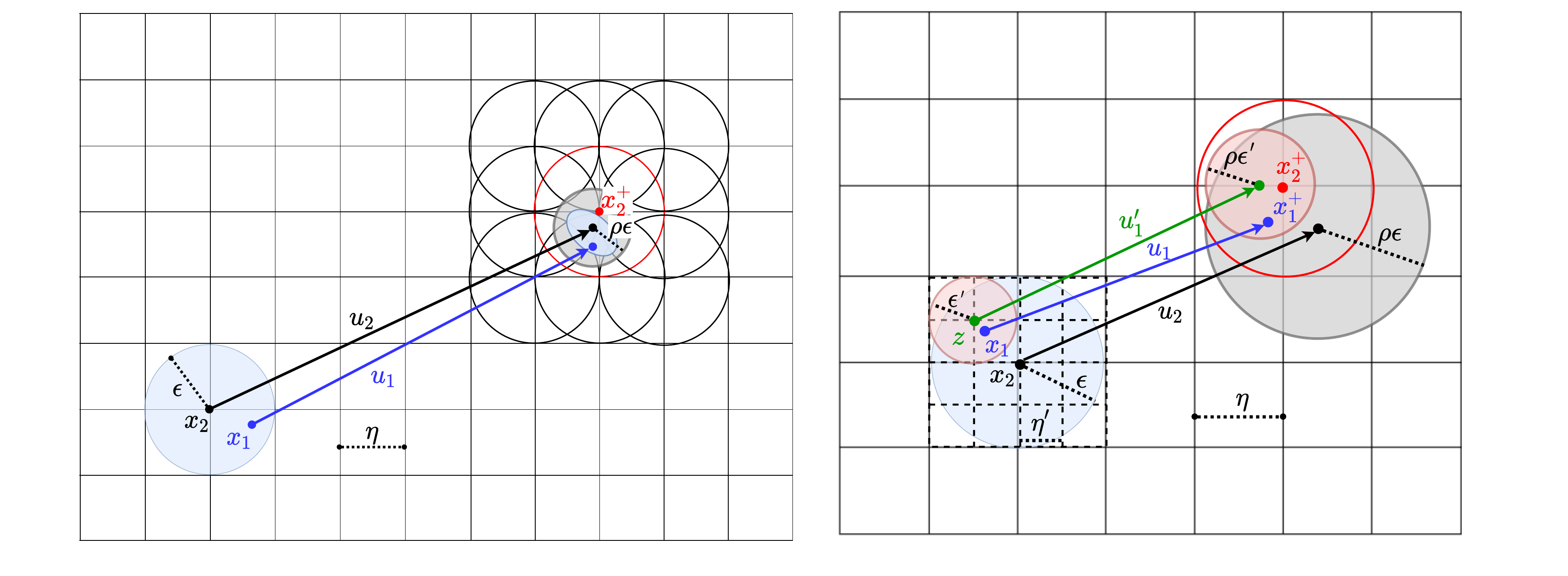}
    \caption{
    Illustration of $(x_2, u_2, x_1, u_1,x_2^+) \in R_e^{\ASRBB}$ (left) and $(x_2, u_2, x_1, u_1) \in R_e^{\ASRB}$ (right) for Example~\ref{example:2D}.
    } 
    \label{fig:Example_ASRBB_ASRB}
\end{figure}

\begin{proposition}\label{prop:example_ASRBB}
Given the additional assumption that $\rho<1$ and $\eta\le  \tfrac{2}{\sqrt{n}} \min\left(\overline{\alpha}^{-1}(\epsilon), \gamma^{-1}((1-\rho) \epsilon)\right)$, the system $\Sys_2^{\ASRBB} = (\set{X}_2, \set{U}_2, F_2^{\ASRBB})$ with $F_2^{\ASRBB}$ satisfying $\forall x_2\in\set{X}_2,\ u_2\in\set{U}_{\Sys_2}(x_2)$:
\begin{equation}\label{eq:example_ASRBB}
    F_2^{\ASRBB}(x_2, u_2) \supseteq \argmin_{x_2'\in \set{X}_2}\ \|f(x_2,u_2)-x_2'\|_{\infty},
\end{equation}
ensures $\Sys_1\preceq_R^{\ASRBB}\Sys_2^{\ASRBB}$. Additionally, one can use the specific implementations  
$h_1^{\ASRBB}(x_2,u_2,x_1,x_2^+) = \{\kappa(x_1,x_2,u_2)\}$ and $h_2^{\ASRBB}(x_2,u_2)= \argmin_{x_2'\in \set{X}_2}$ $\|f(x_2,u_2)-x_2'\|_{\infty}$.
\end{proposition}
\begin{pf}
    Applying Proposition~\ref{prop:bissimulation} with parameters $(\eta,\epsilon,\epsilon)$ and $\rho < 1$, the condition on $\eta$ ensures that 
    $\forall x_2 \in \set{X}_2,\ \forall u_2 \in \set{U}_{\Sys_2}(x_2):\ g_{x_2, u_2}(R^{-1}(x_2)) \subseteq R^{-1}(x_2^+)$ with \( x_2^+ = \argmin_{x_2' \in \set{X}_2}\ \|f(x_2, u_2) - x_2'\|_{\infty} \). Therefore, by~\eqref{eq:example_ASRBB}, \( x_2^+ \in F_2^{\ASRBB}(x_2, u_2) \) and for all \( x_1^+ \in F_1(x_1, u_1) \) with \( u_1 = \kappa(x_1, x_2, u_2) \), we have \( (x_1^+, x_2^+) \in R \). This establishes~\eqref{eq:ASRBB_def}, \(\kappa(x_1, x_2, u_2) \in I_R^{\ASRBB}(x_2, u_2,x_1,x_2^+) \) and \( h_2^{\ASRBB} \) implements Definition~\ref{def:interface_relations_specific}.
    \hfill\hfill\qed
\end{pf}

The choice of $F_2^{\ASRBB}(x_2, u_2) = h_2^{\ASRBB}(x_2,u_2)$ results in a deterministic abstraction as illustrated in Figure~\ref{fig:Example_ASRBB_ASRB} (left).
In this context, constructing $\Sys_2^{\ASRBB}$ involves low computational complexity. Designing an abstract transition only requires computing the trajectory of a single point $f(x_2,u_2)$ and identifying the closest grid point.
However, unlike other approaches, this method requires a strong assumption on the system's stability, specifically incremental stabilizability ($\rho<1$).
In~\citet[Section 6]{calbert2024dionysos}, the authors construct a feedforward abstraction relation $\Sys_2^{\ASRBB}$ for an incrementally stabilizable dynamical system, specifically a DC-DC converter defined in~\citet[Section 4.2]{rungger2016scots}, and illustrate the speed-up in the construction of the abstraction resulting from this efficient implementation of $h_2^{\ASRBB}$.


We now construct a finer grid \([\eta' \Z^n]\) with \(0 < \eta' < \eta\); see Figure~\ref{fig:Example_ASRBB_ASRB} (right). Given \(\epsilon' > 0\), we choose \(\eta'\) sufficiently small, satisfying \(\eta' \le \tfrac{2}{\sqrt{n}} \, \overline{\alpha}^{-1}(\epsilon')\), so that every state \(x_1 \in \set{X}_1\) lies within an \(\epsilon'\)-sublevel set \(\set{S}(z, \epsilon')\) of \(V\), centered at some vertex \(z \in [\eta' \Z^n]\) of the refined lattice (see Proposition~\ref{prop:strict_relation}).
Given $x_2\in\set{X}_2$, we denote $\set{Z}_{\eta,\eta',\epsilon,\epsilon'}(x_2)\subseteq [\eta'\Z^n]$ a set of minimal cardinality s.t.
\begin{equation}\label{eq:sub_grid}
    R^{-1}(x_2) \subseteq \bigcup_{z\in \set{Z}_{\eta,\eta',\epsilon,\epsilon'}(x_2)}\set{S}(z, \epsilon').
\end{equation}

\begin{proposition}\label{prop:example_ASRB}
Let $\set{Z}_{\eta,\eta',\epsilon,\epsilon'}(x_2)$ defined in~\eqref{eq:sub_grid} with parameters satisfying
\begin{empheq}[left=\empheqlbrace]{align}
        \eta'& \le \tfrac{2}{\sqrt{n}}\overline{\alpha}^{-1}(\epsilon'), \label{eq:param_1}\\
        \rho \epsilon' &< \epsilon, \label{eq:param_2}\\
        \eta &\le  \tfrac{2}{\sqrt{n}} \min\left(\overline{\alpha}^{-1}(\epsilon), \gamma^{-1}(\epsilon-\rho \epsilon')\right).\label{eq:param_3} 
\end{empheq}
The system $\Sys_2^{\ASRB}= (\set{X}_2,\set{U}_2,F_2^{\ASRB})$ with $F_2^{\ASRB}$ satisfying $\forall x_2\in\set{X}_2,\ u_2\in\set{U}_{\Sys_2}(x_2)$:
\begin{multline*}
F_2^{\ASRB}(x_2,u_2)\supseteq\\ \bigcup_{z\in \set{Z}_{\eta,\eta',\epsilon,\epsilon'}(x_2)} \argmin_{x_2'\in \set{X}_2}\ \|f(z,\kappa(z, x_2, u_2))-x_2'\|_{\infty},
\end{multline*}
ensures $\Sys_1\preceq_R^{\ASRB}\Sys_2^{\ASRB}$.
Additionally, one can use the specific implementations 
\begin{align*}
    h_1^{\ASRB}(x_2,u_2,x_1) &= \{\kappa(x_1,z,u_1')\}\\
    h_2^{\ASRB}(x_2,u_2,x_1,u_1) &= \argmin_{x_2'\in \set{X}_2}\ \|f(z,u_1')-x_2'\|_{\infty},
\end{align*}
where $z = \argmin_{z'\in\set{Z}_{\eta,\eta',\epsilon,\epsilon'}(x_2)} \ \|x_1-z'\|_{\infty}$ and $u_1' = \kappa(z,x_2,u_2)$.
\end{proposition}
\begin{pf}
Let $(x_1, x_2) \in R$ and $u_2 \in \set{U}_{\Sys_2}(x_2)$. By applying Proposition~\ref{prop:strict_relation} with parameters $(\eta', \epsilon')$ satisfying~\eqref{eq:param_1}, we have $x_1 \in \set{S}(z, \epsilon')$ with $z = \argmin_{z' \in \set{Z}_{\eta, \eta', \epsilon, \epsilon'}(x_2)} \|x_1 - z'\|_{\infty}$.
Next, applying Proposition~\ref{prop:bissimulation} with parameters $(\eta, \epsilon', \epsilon)$ satisfying~\eqref{eq:param_2} and \eqref{eq:param_3}, we ensure that $x_1^+ = f(x_1, u_1) \in \set{S}(x_2^+, \epsilon)$ with $u_1 = \kappa(x_1, z, u_1')$, $x_2^+ = \argmin_{x_2' \in \set{X}_2} \|f(z, u_1') - x_2'\|_{\infty}$, and $u_1' = \kappa(z, x_2, u_2)$.
Thus, $x_1^+ \in R^{-1}(x_2^+)$ and by definition of $ F_2^{\ASRB}$,
$x_2^+ \in F_2^{\ASRB}(x_2, u_2)$, which establishes~\eqref{eq:ASRB_def}. Thus, $u_1 \in I_R^{\ASRB}(x_2, u_2, x_1)$ and $h_2^{\ASRB}$ satisfies Definition~\ref{def:interface_relations_specific}.
\hfill\hfill\qed
\end{pf}
As discussed in Section~\ref{sec:discussion_corrective_vs_predictive}, it is possible to predict the next abstract state $x_2^+$ using $h_2^{\ASRB}(x_2, u_2, x_1, u_1)$ before applying the concrete input $u_1$ given by $h_1^{\ASRB}(x_2, u_2,x_1)$. Unlike $\ASRBB$, the $\ASRB$ relation can be constructed without requiring the incremental stabilizability ($\rho<1$).
\section{Conclusion}\label{sec:Conclusion}
Abstraction-based control relies on two core elements: 
the \emph{relation}, ensuring consistent transitions between concrete and abstract states, 
and the \emph{concretization procedure}, which derives a valid controller for the original system from the abstract one.

In this article,  we proposed a systematic approach to characterize simulation relations that refine the $\ASR$, through a concretization procedure featuring a plug-and-play control architecture. To this end, we introduce a universal system transformation, called the \emph{interfaced system}, which is universal in the sense that it is always in $\FRR$ with the abstract system. This allowed us to leverage the necessary and sufficient characterization of $\FRR$ in terms of concretization procedure (Theorem~\ref{th:FRR_concretization}) to a broader class of simulation relations. Specifically, we demonstrated (Corollary~\ref{cor:concrete_controller_augmented}) that if the concrete system is related to the abstraction via a simulation relation, then the abstract controller can be seamlessly connected to the original system through the corresponding interface, irrespective of the specific specification to be enforced, and vice versa. 
This generalizes previous results and provides a unifying framework for systematically characterizing simulation relations. 
Finally, we discussed the pros and cons of different relations using a common example.

Our future research will focus on three main directions. 
First, while we have illustrated our approach using five key relations relevant for the design of smart abstractions, we plan to develop a systematic procedure to identify the interface characterizing any refined notion of the alternating simulation relation.
Second, while the objective of this work is to highlight the benefits of simpler or more structured control architectures enabled by conservative simulation relations refining $\ASR$, we aim to investigate which classes of systems allow these relations to be practically constructed.
Section~\ref{sec:Constructions} provides a first step in this direction by showing that relations such as \(\ASR\), \(\MCR\), and \(\ASRB\) can be built for arbitrary systems, whereas \(\ASRBB\) additionally requires incremental stabilizability.
Third, we aim to extend our current framework, which focuses on state-feedback control of simple systems, to \emph{output feedback} control of systems with output maps. Various types of relations for such systems are well-established in the literature~\citep{majumdar2020abstraction,apaza2020symbolic,schmuck2014asynchronous,coppola2022data}, and we plan to classify and characterize them in terms of their associated control architectures.

\begin{ack}
JC is a FRIA Research Fellow. 
RJ is a FNRS honorary Research Associate. This project has received funding from the European Research Council (ERC) under the \emph{European Union's Horizon 2020 research and innovation programme} under grant agreement No 864017 - L2C, from the Horizon Europe programme under grant agreement No101177842 - Unimaas, and from the ARC (French Community of Belgium)-SIDDARTA.
\end{ack}

\bibliographystyle{agsm} 
\bibliography{main.bib}  
\appendix


\section{Proof of Theorem~\ref{th:augmented_system}}\label{pf:augmented_system}

   We start by proving (1). From the definition of \( I_R^{\T} \) in~\eqref{eq:interface_abstract_concrete_input}, it follows that the tuple \((\set{Z}_1, h_1^{\T}, h_2^{\T}, \widetilde{R})\) in Definition~\ref{def:interface_relations_specific} is an interface from \(\Sys_2\) to \(\Sys_1\). Finally, we show that \(\widetilde{R}\) in~\eqref{eq:R_to_Rtilde} satisfies Definition~\ref{as:ASRB}. Let \(x_1, x_1'\) be such that \(\widetilde{R}((x_1, x_2)) \ne \myemptyset\) and \(\widetilde{R}((x_1', x_2)) \ne \myemptyset\). Then we have \(\widetilde{R}((x_1, x_2)) \cap \widetilde{R}((x_1', x_2)) = \{x_2\}\ne \myemptyset\).
   We now turn to proving (2). By Theorem~\ref{th:concrete_controller_augmented}, we have \(\widetilde{\Sys}_1^{\T} \preceq_{\widetilde{R}}^{\FRR} \Sys_2\). From Definition~\ref{def:FRR}, it follows that for any \(((x_1, z_1), x_2) \in \widetilde{R}\)
    \begin{itemize}
        \item[$(i)$] $\set{U}_{\Sys_2}(x_2)\subseteq \set{U}_{\widetilde{\Sys}_1^{\T}}((x_1,z_1))$;
        \item[$(ii)$] $\forall u_2\in\set{U}_{\Sys_2}(x_2)\ \forall (x_1^+,z_1^+)\in \widetilde{F}_1((x_1,z_1), u_2):\ \widetilde{R}((x_1^+,z_1^+))\ne \myemptyset$ and $\widetilde{R}((x_1^+,z_1^+))\subseteq F_2(x_2, u_2)$.
    \end{itemize}
    
    Let $(x_1,x_2)\in R$ and $u_2\in \set{U}_{\Sys_2}(x_2)$. This implies the existence of $z_1$ such that $((x_1,z_1),x_2)\in\widetilde{R}$. By condition $(i)$, we have $u_2\in \set{U}_{\widetilde{\Sys}_1}((x_1,z_1))$, meaning $\widetilde{F}_1((x_1,z_1),u_2)\ne \myemptyset$. Thus, by Definition~\ref{def:augmented_system}, for any $u_1\in h_1^{\T}(\nu_1)$, $u_1\in \set{U}_{\Sys_1}(x_1)$, and for all $x_1^+ \in F_1(x_1,u_1)$, $z_1^+ \in h_2^{\T}(\nu_2)$, it holds that $(x_1^+,z_1^+)\in\widetilde{F}_1((x_1,z_1),u_2)$.
    In addition, by $(ii)$, $\widetilde{R}((x_1^+,z_1^+))\ne \myemptyset$ and $\widetilde{R}((x_1^+,z_1^+))\subseteq F_2(x_2, u_2)$.
    
    The rest of the proof is specific to each relation type:
    \begin{itemize}[left=0.02cm, labelsep=0.01cm]
        \item $\ASR$:
        According to Lemma~\ref{lem:relation_inclusion}, $\widetilde{R}((x_1^+,z_1^+)) \subseteq R(x_1^+)$.
        Therefore, 
        $F_2(x_2, u_2) \cap R(x_1^+) \supseteq \widetilde{R}((x_1^+,z_1^+)) \ne\myemptyset,$
        which is equivalent to~\eqref{eq:ASR_def}.  
            
        \item $\ASRB$:
        We consider $u_1\in h_1^{\ASRB}(z_1,u_2,x_1)$ and $z_1^+\in h_2^{\ASRB}(z_1,u_2,u_1,x_1)$. 
        Since $\widetilde{R}$ satisfies Definition~\ref{as:ASRB}, it holds that
        \[
        \set{A} \defEqual \left(\bigcap_{x_1^+\in F_1(x_1, u_1)}\ \widetilde{R}((x_1^+,z_1^+))\right)\ne \myemptyset,
        \]
        as $\widetilde{R}((x_1^+,z_1^+))\ne \myemptyset$ by $(ii)$.
        Let $x_2^+\in \set{A}$. Firstly, $x_2^+\in F_2(x_2,u_2)$ since $\widetilde{R}((x_1^+,z_1^+))\subseteq F_2(x_2, u_2)$ by $(ii)$. Secondly, according to Lemma~\ref{lem:relation_inclusion}, $\widetilde{R}((x_1^+,z_1^+))\subseteq R(x_1^+)$, therefore $x_2^+\in R(x_1^+)$, which is equivalent to~\eqref{eq:ASRB_def}.
    
        \item $\ASRBB$: 
        Let $z_1^+\in h_2^{\ASRBB}(z_1,u_2)$. Since $\widetilde{R}$ satisfies Definition~\ref{as:ASRB}, it holds that
        {\footnotesize \[
        \set{A} \defEqual
        \bigcap_{\begin{aligned}
        u_1&\in h_1(z_1,u_2,x_1,z_1^+)\\
        x_1^+&\in F_1(x_1, u_1)
        \end{aligned}} \ \widetilde{R}((x_1^+,z_1^+)),
        \]}
        as $\widetilde{R}((x_1^+,z_1^+))\ne \myemptyset$ by $(ii)$.
        Let $x_2^+\in \set{A}$. Firstly, $x_2^+\in F_2(x_2,u_2)$ since $\widetilde{R}((x_1^+,z_1^+))\subseteq F_2(x_2, u_2)$ by $(ii)$. Secondly, according to Lemma~\ref{lem:relation_inclusion}, $\widetilde{R}((x_1^+,z_1^+))\subseteq R(x_1^+)$, therefore $x_2^+\in R(x_1^+)$, which is equivalent to~\eqref{eq:ASRBB_def}.
    
        \item $\MCR$: 
        By definition of $R$ in~\eqref{eq:Rtilde_to_R}, we have 
        \[
        R(x_1^+) = \bigcup_{z_1^+\in h_2^{\MCR}(x_1^+)} \widetilde{R}((x_1^+,z_1^+)),
        \]
        which establishes that $R(x_1^+) \ne\myemptyset$ and $R(x_1^+)\subseteq F_2(x_2,u_2)$, which is equivalent to~\eqref{eq:MCR_def}.\hfill\hfill\qed
    \end{itemize}

\section{Useful results}
\begin{lemma}\label{lem:equivalence_concrete_traj}
    Let \(\Sys_1\) and \(\Sys_2\) in~\eqref{eq:simple_sys}, and \((\set{Z}_1, h_1, h_2, \widetilde{R})\) an interface from \(\Sys_2\) to \(\Sys_1\) with associated interfaced system \(\widetilde{\Sys}_1\) (Definition~\ref{def:augmented_system}). Then, for any controller \(\Cont_2\) f.c. with \(\Sys_2\), the concretized controller \(\Cont_1\) (Definition~\ref{def:concretized_controller}) and \(\widetilde{\Cont}_1 = \Cont_2 \circ \widetilde{R}\) satisfy
    $\seq{x}_1\in \B(\Cont_1\times \Sys_1) \Leftrightarrow  \seq{\widetilde{x}}_1\in \B(\widetilde{\Cont}_1\times \widetilde{\Sys}_1),$
    where $\seq{x}_1 = \pi_{\set{X}_1}(\seq{\widetilde{x}}_1)$~\eqref{eq:projection}.
\end{lemma}
\begin{pf}
    By Definition~\ref{def:serial_composition}, $\widetilde{\Cont}_1 = \Cont_2 \circ \widetilde{R}= (\set{X}_{\Cont_2}, \set{X}_1 \times \set{Z}_1, \set{V}_{\Cont_2}, \set{U}_2, F_{\Cont_2}, H_{\widetilde{\Cont}_1})$ where \(H_{\widetilde{\Cont}_1}(x_{\Cont_2}, (x_1, z_1)) = H_{\Cont_2}(x_{\Cont_2}, \widetilde{R}(x_1, z_1))\), since \(\widetilde{R}\) is a static system.
    Thus, \(\seq{\widetilde{x}}_1 \in \B(\widetilde{\Cont}_1 \times \widetilde{\Sys}_1)\) if and only if there exist sequences \(\seq{u}_1, \seq{x}_2, \seq{u}_2, \seq{x}_{\Cont_2}, \seq{v}_{\Cont_2}\) satisfying~\eqref{eq:interface_closed_loop_augmented}  with \(\seq{x}_1 = \pi_{\set{X}_1}(\seq{\widetilde{x}}_1)\) and \(\seq{z}_1 = \pi_{\set{Z}_1}(\seq{\widetilde{x}}_1)\). This establishes that \(\seq{x}_1 \in \B(\Cont_1 \times \Sys_1)\).
    \hfill\hfill\qed
\end{pf}

\begin{lemma}\label{lem:relation_inclusion}
    Let $\widetilde{R}\subseteq (\set{X}_1\times \set{Z}_1)\times \set{X}_2$, and $R\subseteq \set{X}_1\times \set{X}_2$ in~\eqref{eq:Rtilde_to_R}. 
    If $\widetilde{R}((x_1,z_1))\ne \myemptyset$, then $\widetilde{R}((x_1,z_1)) \subseteq R(x_1)$.
\end{lemma}
\begin{pf} 
By definition, 
\[
R(x_1) = \cup_{\smash{(x_1, z_1) \mid \widetilde{R}((x_1, z_1)) \neq \myemptyset}} \ \widetilde{R}((x_1, z_1)).\tag*{\qed}
\]
\end{pf}

\begin{proposition}\label{prop:strict_relation}
Let $V$ be a growth-bound function as in Definition~\ref{def:Lyap} and parameters $(\eta,\epsilon)>0$.
If $\eta\le \tfrac{2}{\sqrt{n}}\overline{\alpha}^{-1}(\epsilon)$, then for all $x\in\R^n$, $\ x\in \mathcal{S}(x_2,\epsilon)$ with $x_2 = \argmin_{c\in [\eta \Z^n]} \ \|x-c\|_{\infty}$.
\end{proposition}
\begin{pf}
    Let $x\in\R^n$ and $x_2 = \argmin_{c\in [\eta \Z^n]} \ \|x-c\|_{\infty}$. Then $\|x-x_2\|_{\infty}\le \tfrac{\eta}{2}$, and we have $V(x,x_2)\le \overline{\alpha}(\|x-x_2\|_2)\le \overline{\alpha}(\sqrt{n}\|x-x_2\|_{\infty})\le \overline{\alpha}\left(\sqrt{n}\tfrac{\eta}{2}\right)\le \epsilon,$
    because  $\overline{\alpha}$ is a $\mathcal{K}$ function. \hfill\hfill\qed
\end{pf}

\begin{proposition}\label{prop:over_approx}
    Let $V$ be a growth-bound function as in Definition~\ref{def:Lyap}. Given $x_2 \in \R^n$, $u_2 \in \R^{\nu}$, and $\epsilon > 0$. Then, the following holds $g_{x_2,u_2}(\set{S}(x_2,\epsilon)) \subseteq \set{S}(f(x_2,u_2),\rho \epsilon),$
    where $g_{x_2,u_2}(x) = f(x, \kappa(x, x_2, u_2))$.
\end{proposition}
\begin{pf}
    Let $x\in \set{S}(x_2,\epsilon)$.
    By~\eqref{eq:lyapunov}, it holds $$V(f(x,\kappa(x,x_2,u_2)), f(x_2,u_2))\le \rho V(x,x_2)\le \rho \epsilon,$$
    which involves $g_{x_2,u_2}(x)\in \set{S}(f(x_2,u_2),\rho \epsilon)$.
    \hfill\hfill\qed
\end{pf}

\begin{proposition}[{\citet[Eq. (9)]{girard2009approximately}}]\label{prop:bissimulation}
    Let $V$ be a growth-bound function as in Definition~\ref{def:Lyap} satisfying~\eqref{eq:as_sup} and parameters $(\eta,\epsilon,\epsilon')>0$.
    If $\epsilon'-\rho\epsilon> 0$, then the choice $\eta \le \tfrac{2}{\sqrt{n}}\gamma^{-1}(\epsilon'-\rho\epsilon)$ ensures that
    $$\forall x_2\in\R^n\ \forall u_2\in\R^{\nu}:\ 
    g_{x_2,u_2}(\mathcal{S}(x_2, \epsilon)) \subseteq \mathcal{S}(x_2', \epsilon'),$$
    with $x_2' = \argmin_{x'\in[\eta \Z^n]}\  \|x'-f(x_2, u_2)\|_{\infty}$ and $g_{x_2,u_2}(x) = f(x, \kappa(x, x_2, u_2))$.
    Special case: when $V(x,y)=\|x-y\|_2$ and $\epsilon=\epsilon'$, the condition on $\eta$ reduces to $\eta\le\tfrac{2}{\sqrt{n}}\epsilon (1-\rho)$.
\end{proposition}
\begin{pf}
    Let $x_1\in \mathcal{S}(x_2,\epsilon)$, $x_1^+ = f(x_1,\kappa(x_1,x_2,u_2))$, $x_2^+=f(x_2, u_2)$ and $x_2' = \argmin_{x'\in[\eta \Z^n]}\  \|x'-x_2^+\|_{\infty}$. Then, $\|x_2' - x_2^+\|_{\infty} \leq \tfrac{\eta}{2}$.
    Using~\eqref{eq:as_sup} and~\eqref{eq:lyapunov}, and since $\gamma$ is a $\mathcal{K}_{\infty}$ function, we have
    \begin{align*}
        V(x_1^+,x_2') 
        &\le V(x_1^+, x_2^+) + \gamma(\|x_2'-x_2^+\|_2)\\
        &\le \rho V(x_1, x_2) + \gamma(\sqrt{n}\|x_2'-x_2^+\|_{\infty})\\
        &\le \rho \epsilon + \gamma\left(\sqrt{n}\tfrac{\eta}{2}\right)\\
        &\le \epsilon'
    \end{align*}
    which implies that $x_1^+\in \mathcal{S}(x_2',\epsilon')$.
    \hfill\hfill\qed
\end{pf}


\small

\setlength{\columnsep}{13pt}
\begin{wrapfigure}{l}{0.12\textwidth}
\vspace{-15pt}
\includegraphics[
    width=\linewidth,
    trim={5 50 10 5}, 
    clip
]{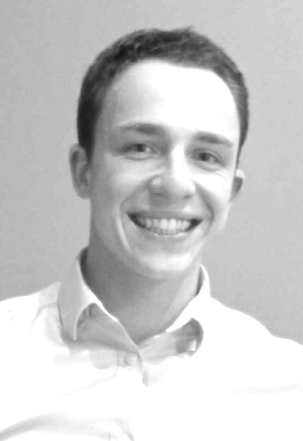}
\vspace{-15pt}
\end{wrapfigure}
\textbf{Julien Calbert} received the B.S. degree in computer engineering from UCLouvain, Belgium, in 2018, and the M.Eng. degree in applied mathematics from UCLouvain in 2020. He obtained the Ph.D. degree in applied mathematics from UCLouvain in 2024, as a FRIA fellow (F.R.S.-FNRS), within the ICTEAM Institute under the supervision of Prof. Raphaël M. Jungers. His research interests include verification and control of dynamical systems, control of hybrid systems, optimal control, and symbolic control.

\vspace{0.3cm}

\begin{wrapfigure}{l}{0.12\textwidth}
\vspace{-10pt}
\includegraphics[width=\linewidth]{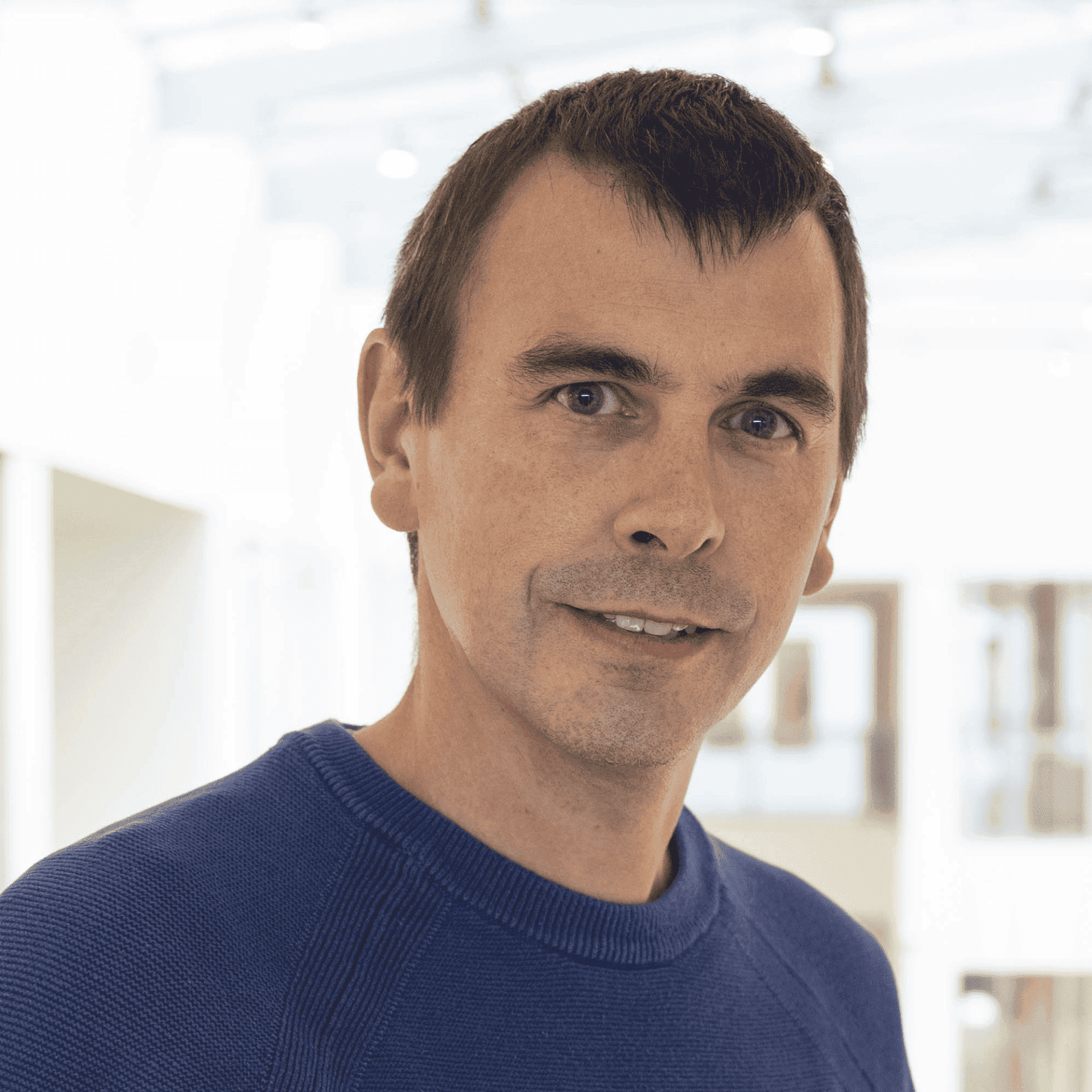}
\vspace{-10pt}
\end{wrapfigure}
\textbf{Antoine Girard} is a Senior Researcher at CNRS and a member of the Laboratory of Signals and Systems. He is also an Adjunct Professor at CentraleSupélec, Université Paris-Saclay. He received the Ph.D. degree from Grenoble Institute of Technology, in 2004. From 2004 to 2006, he held postdoctoral positions at University of Pennsylvania and Université Grenoble-Alpes. From 2006 to 2015, he was an Assistant/Associate Professor at the Université Grenoble-Alpes. His main research interests deal with analysis and control of hybrid systems with an emphasis on computational approaches, formal methods and applications to cyber-physical and autonomous systems. Antoine Girard is an IEEE Fellow. In 2015, he was appointed as a junior member of the Institut Universitaire de France (IUF). In 2016, he was awarded an ERC Consolidator Grant. He received the George S. Axelby Outstanding Paper Award from the IEEE Control Systems Society in 2009, the CNRS Bronze Medal in 2014, and the European Control Award in 2018.

\vspace{0.3cm}

\begin{wrapfigure}{l}{0.12\textwidth}
\vspace{-10pt}
\includegraphics[width=\linewidth]{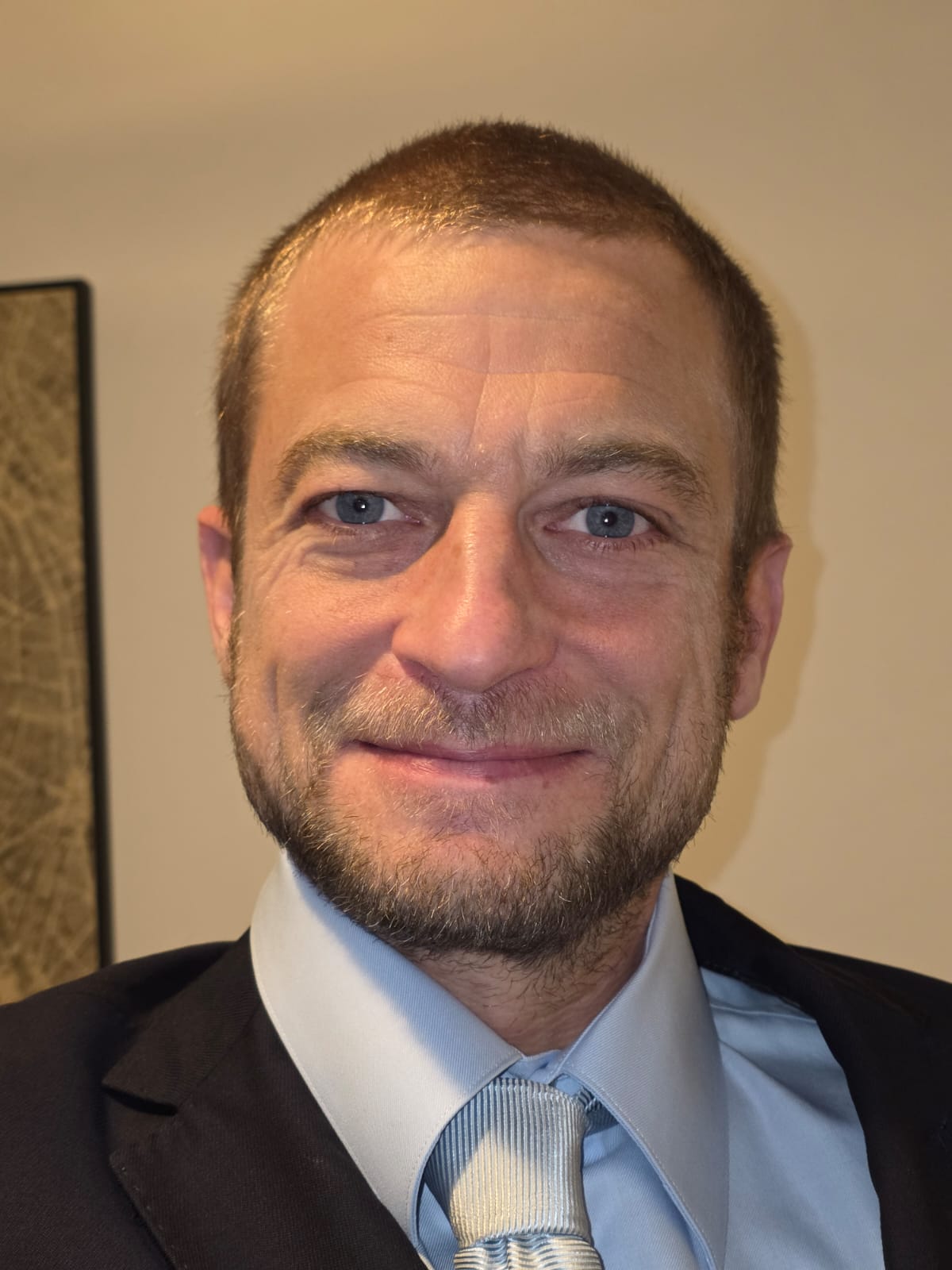}
\vspace{-10pt}
\end{wrapfigure}
\textbf{Raphaël M. Jungers} is a Professor at UCLouvain, Belgium. His main interests lie in the fields of Computer Science, Graph Theory, Optimization and Control. He received a Ph.D. in Mathematical Engineering from UCLouvain (2008), and a M.Sc. in Applied Mathematics, both from the Ecole Centrale Paris, (2004), and from UCLouvain (2005). 
He has held various invited positions, at the Université Libre de Bruxelles (2008-2009), at the Laboratory for Information and Decision Systems of the Massachusetts Institute of Technology (2009-2010), at the University of L´Aquila (2011, 2013, 2016), at the University of California Los Angeles (2016-2017), and at Oxford University (2022-2023). He has been serving as Vice President of IEEE CSS, in charge of Conference Activities (2026-2027). 
He is a FNRS, BAEF, and Fulbright fellow. He has been an Editor at large for the IEEE CDC, Associate Editor for the IEEE CSS Conference Editorial Board, and the journals NAHS (2015-2016), Systems and Control Letters (2016-2017), IEEE Transactions on Automatic Control (2015-2020), Automatica (2020-2025). He is currently serving as deputy Editor in Chief for NAHS. He was the recipient of the IBM Belgium 2009 award and a finalist of the ERCIM Cor Baayen award 2011. He was the co-recipient of the SICON best paper award 2013-2014, the HSCC2020 best paper award, the NAHS 2020-2022 best paper award, and he is the recipient of an ERC 2019 award. He is an IEEE Fellow (class 2025).
\end{document}